\documentclass[epj]{webofc}
\usepackage[utf8]{inputenc}
\usepackage[varg]{txfonts}
\usepackage{booktabs}
\usepackage{xcolor}
\usepackage{braket, bm}
\definecolor{darkred}{rgb}{0.4,0.0,0.0}
\definecolor{darkgreen}{rgb}{0.0,0.4,0.0}
\definecolor{darkblue}{rgb}{0.0,0.0,0.4}
\usepackage[bookmarks,linktocpage,colorlinks,
    linkcolor = darkred,
    urlcolor  = darkblue,
    citecolor = darkgreen]{hyperref}
\usepackage{subfigure}
\wocname{EPJ Web of Conferences}
\woctitle{Lattice2017}
\begin{document}
%
\selectlanguage{english}
\title{%
Lattice QCD input for nuclear structure and reactions
}
\author{%
\firstname{Zohreh} \lastname{Davoudi}\inst{1}\fnsep\thanks{Affiliation at the time of the conference: 
Center for Theoretical Physics, Massachusetts Institute of Technology, Cambridge, MA 02139, USA.}\fnsep\thanks{\email{davoudi@umd.edu}}
}
\institute{%
Department of Physics and Maryland Center for Fundamental Physics
\\
University of Maryland, College Park, MD 20742, USA
}
\abstract{%
 Explorations of the properties of light nuclear systems beyond their lowest-lying spectra have begun with Lattice Quantum Chromodynamics. While progress has been made in the past year in pursuing calculations with physical quark masses, studies of the simplest nuclear matrix elements and nuclear reactions at heavier quark masses have been conducted, and several interesting results have been obtained. A community effort has been devoted to investigate the impact of such  Quantum Chromodynamics input on the nuclear many-body calculations. Systems involving hyperons and their interactions have been the focus of intense investigations in the field, with new results and deeper insights emerging. While the validity of some of the previous multi-nucleon studies has been questioned during the past year, controversy remains as whether such concerns are relevant to a given result. In an effort to summarize the newest developments in the field, this talk will touch on most of these topics.
}
\maketitle

\section{Introduction}\label{sec:intro}
\noindent
There are at least four (interconnected) frontiers in nuclear physics research that have been driven by experimental, theoretical and computational advances in recent years. These are aimed at enhancing our understanding of ordinary matter in nature, that described by the Standard Model (SM) of Particle Physics, and establishing hints of the existence of mechanisms beyond the SM. One frontier, on the theoretical side, relies on \emph{ab initio} quantum many-body calculations and precise knowledge of nuclear and hypernuclear forces, in order to advance studies of dense matter and neutron-rich environments, and to support experiments such as that planned at the Facility for Rare Isotopes Beams (FRIB) in the U.S.\footnote{The sole mention of the experimental programs in the U.S. in this talk is to provide examples. This choice, by no means, is meant to undermine the strong experimental effort in hadronic and nuclear physics worldwide.} In another frontier, the goal is to provide precise constraints on some of the most fundamental nuclear reaction cross sections in nature, such as those occurring in the $pp$ chain in Sun, and those relevant for  energy-production facilities such as the National Ignition Facility (NIF), U.S. Hadronic and nuclear structure, as considered from the perspective of the SM, is the subject of an extensive experimental effort in the field. The electron-Ion Collider in the U.S., if built, can provide unique perspectives on the role of quarks and gluons in hadrons and nuclei. To confront such findings with theoretical expectations requires first-principles calculations of structure properties of nucleon and bound nuclei. Last but not least, a large community effort is dedicated to discovering violations of the fundamental symmetries of nature, such as the time-reversal invariance (through searches for the electric dipole moment of atoms and nuclei) and lepton-number conservation (through searches for the neutrinoless double-beta decay of various isotopes). Additionally, there are ongoing direct dark-matter detection experiments which use heavy isotopes as targets. Clearly, to establish that new mechanism are needed to explain any experimental findings, theoretical expectations for SM contributions must reach a level of precision that is comparable with that of experiment, a challenging goal given the complexity of the nuclei involved.

The only reliable first-principles method, whose degrees of freedom are the quarks and gluons of the SM, and whose only input parameters are those of the strong interactions, is lattice quantum chromodynamics (LQCD). Whether to constrain forces between nucleons or the reaction cross sections among them to compensate for the scarcity of experimental data, or to provide SM contributions to processes that are sensitive to physics beyond the SM in order to complement experiments, LQCD input for multi-nucleon observables is needed more than ever. Although the field of LQCD for nuclear physics is still in its infancy, primarily due to the complexity of the calculations and the magnitude of computational resources involved, the goals of the program appear to be within the reach in the upcoming years, in particular with the emergence of \emph{Exascale} high-performance computing~\cite{ExascaleNP}.

The first LQCD study of light nuclei and hypernuclei was conducted in 2012 by the NPLQCD collaboration~\cite{Beane:2012vq}. This study obtained the lowest-lying spectra of systems with atomic number $A \leq 4$, at a $SU(3)$ flavor-symmetric point, with quark masses that correspond to a mass of approximately $806~\tt{MeV}$ for the pions and kaons. Quantum Electrodynamics (QED) interactions were not accounted for in this study, a statement that still holds for all multi-nucleon calculations performed to date. It is known how to implement QED in a LQCD calculation, with strategies to alleviate the large finite-volume effects arising from an infinite-range force~\cite{Hayakawa:2008an, Davoudi:2014qua, Borsanyi:2014jba, Endres:2015vpi, Lucini:2015hfa}. Impressive progress has been made in constraining the QED contributions to the mass splittings between the members of meson and baryon multiplets (e.g., Refs.~\cite{Borsanyi:2014jba, Horsley:2015eaa, Patella:2017fgk, Boyle:2017gzv}). Additionally, there have been frameworks suggested to obtain a wider range of observables from lattice QED+QCD calculations, among which are hadron-hadron elastic scattering~\cite{Beane:2014qha}, decay matrix elements involving charged initial and final states (e.g., $\pi^+ \to \mu^+ \nu_{\mu}$)~\cite{Carrasco:2015xwa, Lubicz:2016xro}, and the isospin breaking contributions to the anomalous magnetic moment of the muon~\cite{Boyle:2017iqy, Giusti:2017ier, Bussone:2017xkb, Boyle:2017gzv}. It is therefore conceivable that such advances will enter the studies of multi-nucleon systems in short term.

Following the first calculation of nuclei, further studies emerged by several collaborations, in particular in an effort to lower the value of the light quark masses towards the physical values (see e.g.,~\cite{Yamazaki:2012hi, Orginos:2015aya, Yamazaki:2015asa}). Additionally, nucleon-nucleon, nucleon-hyperon and hyperon-hyperon scattering at low-energies have been studied at multiple values of the quark masses and some of the implications for such systems in nature, in particular in the H-dibaryon channel, have been explored~\cite{Beane:2006gf,Nemura:2008sp,Beane:2009py,Beane:2010hg,Beane:2011zpa,Beane:2011iw,Inoue:2011ai,Beane:2012ey,Yamada:2015cra,Berkowitz:2015eaa, Miyamoto:2017tjs, Gongyo:2017fjb}. Progress in studies of nuclear and hypernuclear interactions were reported in this conference, and are summarized in Sec.~\ref{sec:BB}. Based on methodologies used, these studies can be divided to two categories. Unfortunately, conclusions reached regarding the lowest-lying states of few-nucleon and hyperon systems at similar values of the quark masses have not been in agreement between the two categories. Most groups rely on the ground-state saturation in the nuclear correlation functions to obtain the lowest-lying finite-volume energy spectra. In the two-body sector, with sufficiently large volumes such that exponential corrections due to the range of hadronic interactions are negligible, these energies can be used to obtain elastic scattering amplitudes at the corresponding energies with the use of L\"uscher's method~\cite{Luscher:1986pf, Luscher:1990ux}. Additionally, there has been ongoing progress in finding extensions of L\"uscher's method or its asymptotic form to more than two hadrons~\cite{Polejaeva:2012ut, Briceno:2012rv, Hansen:2014eka, Briceno:2017tce, Hammer:2017uqm, Hammer:2017kms, Detmold:2008gh, Hansen:2016fzj, Meissner:2014dea, Konig:2017krd, Mai:2017bge}. On the other hand, the HAL~QCD collaboration uses an incarnation of the ``potential'' method that avoids the need for ground-state saturation in multi-nucleon correlation functions~\cite{Ishii:2006ec, Aoki:2008hh, Aoki:2009ji, HALQCD:2012aa, Miyamoto:2017tjs, Gongyo:2017fjb}. Extracted potentials, along with LQCD Bethe-Saltpeter wavefunctions, are used to obtain the scattering phase shifts. Extensive criticism of the potential method has appeared in literature in recent years~\cite{Beane:2010em, Walker-Loud:2014iea, Yamazaki:2015nka, Savage:2016egr, Yamazaki:2017gjl}, attributing the conflict between the existing results on multi-nucleon systems to uncontrolled systematic uncertainties inherent in the potential method, a suspicion that seems to have been confirmed by the HAL~QCD collaboration through their investigations of the $\pi\pi$ scattering in the $\rho$ channel, as reported in this conference~\cite{Kawai}. On the other hand, within the past year, HAL~QCD presented a series of criticisms of all the work in literature based on a ground-state saturation in multi-nucleon systems~\cite{Iritani:2016jie, Iritani:2017rlk} (see also the contributions to this conference in Refs.~\cite{Iritani:2017wvu, Aoki:2017byw}). Rebuttals of these criticisms by the NPLQCD collaboration and the PACS-CS collaboration also appeared within the last year, putting doubts on the HAL~QCD's investigations of their work~\cite{Beane:2017edf, Wagman:2017tmp, Yamazaki:2017euu, Yamazaki:2017jfh}. Sec.~\ref{sec:robustness} summarizes some of these criticisms and the arguments presented in their response.

With the confidence in the ground-state spectra of multi-nucleon systems at heavier values of the quark masses, NPLQCD collaboration has extended the exploration of properties of light nuclei to their structure properties and some of their simplest reactions. Work that was presented in previous conferences included a LQCD determination of the magnetic moment (and isovector) magnetic polarizabilities of nucleons and light nuclei with $A \leq 3$~\cite{Beane:2014ora, Chang:2015qxa, Savage:2016egr, Parreno:2016fwu, Parreno:2017vbw}, as well as the first LQCD constraint on a reaction matrix element for the $np \to d \gamma$ radiative capture process at heavy quark masses corresponding to pion masses of $m_{\pi} \approx 806~\tt{MeV}$ and $m_{\pi} \approx 450~\tt{MeV}$~\cite{Beane:2015yha}. Further progress was made by the collaboration in the past year by extending such calculations to axial properties of light nuclei at a pion mass of $\approx 806~\tt{MeV}$, obtaining the cross section for the $pp$ fusion process, $pp \to d e^+ \nu_e$~\cite{Savage:2016kon}, and the matrix element relevant for the neutrinofull double-$\beta$ decay process, $nn \to ppee \overline{\nu}_e \overline{\nu}_e $~\cite{Shanahan:2017bgi, Tiburzi:2017iux}. These results will be summarized in Sec.~\ref{sec:reactions}. Additionally, the Gamov-Teller matrix element contributing to the tritium $\beta$ decay is obtained from a constraint on the axial charge of $^3$He~\cite{Savage:2016kon}. The gluonic probes are being used to get further insight into the structure of nucleons and nuclei, and preliminary results at large values of the quark masses were reported in this conference, as presented in Sec.~\ref{sec:structure}.

I will end this talk in Sec.~\ref{sec:GW-noise} by presenting some of the recent progress in tackling challenges associated with multi-nucleon calculations. A method that has been employed to suppress the excited state contributions in the multi-nucleon correlation functions at early times will be briefly mentioned. Furthermore, a new understanding of the late-time behavior of the noisy nuclear correlation functions will be discussed, along with practical consequences for the extraction of the ground-state energy of hadronic systems obtained from the late-time region of the correlation functions. I summarize and conclude in Sec.~\ref{sec:summary}.

\section{On the robustness of multi-baryon calculations}\label{sec:robustness}
\noindent
The Euclidean nature of LQCD calculations allows a spectral decomposition of the momentum-projected hadronic correlation functions, $C_{\hat{\mathcal{O}},\hat{\mathcal{O}}'}(\tau;\mathbf{d})$, with only a finite number of terms contributing significantly at late times: 
\begin{center}
\begin{eqnarray}
C_{\hat{\mathcal{O}},\hat{\mathcal{O}}'}(\tau;\mathbf{d})=\sum_{\mathbf{x}} e^{2\pi i\mathbf{d} \cdot \mathbf{x} /L} \langle 0 | \hat{\mathcal{O}}'(\mathbf{x},\tau) \hat{\mathcal{O}}^{\dagger}(\mathbf{0},0) | 0\rangle=\mathcal{Z}'_0\mathcal{Z}_0^{\dagger}e^{-E^{(0)}\tau}+\mathcal{Z}'_1\mathcal{Z}_1^{\dagger}e^{-E^{(1)}\tau}+\dots.
\label{eq:corr-funct}
\end{eqnarray}
\end{center}
Here, $\mathbf{d}$ is the boost vector of the system in units of $2\pi/L$, $\mathcal{Z}$ ($\mathcal{Z}'$) denotes the overlap of the interpolating operator $\hat{\mathcal{O}}$ ($\hat{\mathcal{O}}'$) onto the corresponding eigenstates of the system, with subscripts ``$0$'' and ``$1$'' referring to the ground state and the first excited state, respectively. $E^{(0)}$ and $E^{(1)}$ are the ground and first excited-state energies, respectively, and the ellipsis refer to contributions from additional higher-energy states. This correlation function at intermediate to late times can be fit to multi or single-exponential forms using a variety of methods. To enhance the confidence in the extracted ground state, specially with closely-spaced energy levels in the system, means such as variational methods can be employed to orthogonalize the second and higher levels against the ground state. Unfortunately, while such a method is affordable for the case of multi-meson systems, the complexity and a sever signal-to-noise problem in multi-baryon calculations has so far constrained the use of a large basis of interpolating operators required in a variational method. While such methodologies are planned for the next generation of multi-baryon calculations and will be viable with larger computational resources, the question is what can be done with a single or few choices of interpolating operators? If the multi-baryon system in the infinite volume presents a bound state, with a nonnegligible gap to the elastic scattering thresholds, it is conceivable that such a gap persists in the finite volume, and the ground state energy can be extracted confidently from the large-time behavior of the correlation function in Eq. (\ref{eq:corr-funct}). It is also conceivable that with proper choices of interpolating operators, by which a significantly larger overlap to the ground state (or each of the other states) is achieved, the momentum-projected correlation function becomes dominantly a single exponential even at earlier times, during which the signal has not yet completely degraded. These two expectations have been the logical basis of the first LQCD determinations of the lowest-lying spectra of multi-baryon systems such as those reported in Refs.~\cite{Beane:2012vq, Orginos:2015aya} by the NPLQCD collaboration and in Refs.~\cite{Yamazaki:2012hi, Yamazaki:2015asa} by the PACS-CS collaboration.

During the last year, the validity of the work in literature based on the ground-state saturation of correlation functions was questioned by the HAL~QCD collaboration in Refs.~\cite{Iritani:2016jie, Iritani:2017rlk}, a criticism that seems to have mainly formed by overlooking the remarks mentioned above. Here, I summarize three main aspects of the criticism and the subsequent arguments provided by various group, demonstrating the general irrelevance of the concerns expressed by the HAL~QCD collaboration:

\begin{itemize}
\item{By providing the example of a simple model supplemented by ``mock'' data, authors of Refs.~\cite{Iritani:2016jie} conclude that with the energy gap between the ground state and the first excited state being much smaller than other energy scales in the system, one may be deceived by the appearance of fake plateaus in an ``effective-mass plot'' (EMP)\footnote{An EMP is a plot of the quantity $\mathcal{C}_{\hat{\mathcal{O}},\hat{\mathcal{O}}'}(\tau;\mathbf{d}) =  \log\left[ \frac{C_{\hat{\mathcal{O}},\hat{\mathcal{O}}'}(\tau;\mathbf{d})}{C_{\hat{\mathcal{O}},\hat{\mathcal{O}}'}(\tau+1;\mathbf{d})} \right]$ as a function of time and asymptotes to $E^{(0)}$ as $\tau \to \infty$.} at early times, which are essentially the result of a peculiar linear combination of the first two (few) exponential terms in the correlation functions that mimic a single-exponential form. Given that binding energies, and/or the spacing between adjacent finite-volume energies in the multi-baryon systems are only a few $\tt{MeV}$s (compared with the corresponding gap in the single-nucleon case that is $\mathcal{O}(\Lambda_{\rm{QCD}})$), the leading contaminating exponential term in the correlation function will not die off substantially until $t \gtrsim 15~\tt{fm}$, when the signal has completely been dominated by the noise. The authors of Ref.~\cite{Iritani:2016jie} therefore argue that all previous work on multi-baryon systems based on the assumption of a ground-state saturation have reported fake plateaus, as the correlation functions in these works are being fit to a single exponential form at much earlier times that their estimates dictates. 

What is misleading about this conclusion is that while for generic interpolating operators (``sources'' or ``sinks'') an $\mathcal{O}(1)$ overlap to all states in the volume is plausible (thus enforcing the estimates given above), with physically-motivated source and/or sink operators, the exponential degradation of the signal for the ground state can be compensated by a large overlap factor to the ground state, pushing the start of the single-exponential region in the correlation functions to much earlier times than the naive estimates. Close-to-optimal operators are at the heart of the success of LQCD program for nuclear systems, providing a ``Golden Window'' for extraction of ground-state properties of multi-baryon systems, as was noticed and examined extensively in Refs.~\cite{Beane:2009kya, Beane:2009py, Beane:2009gs}. The issue of the choice of the interpolators will be discussed further shortly.

\begin{figure}[t]
\centering
\includegraphics[scale=0.5]{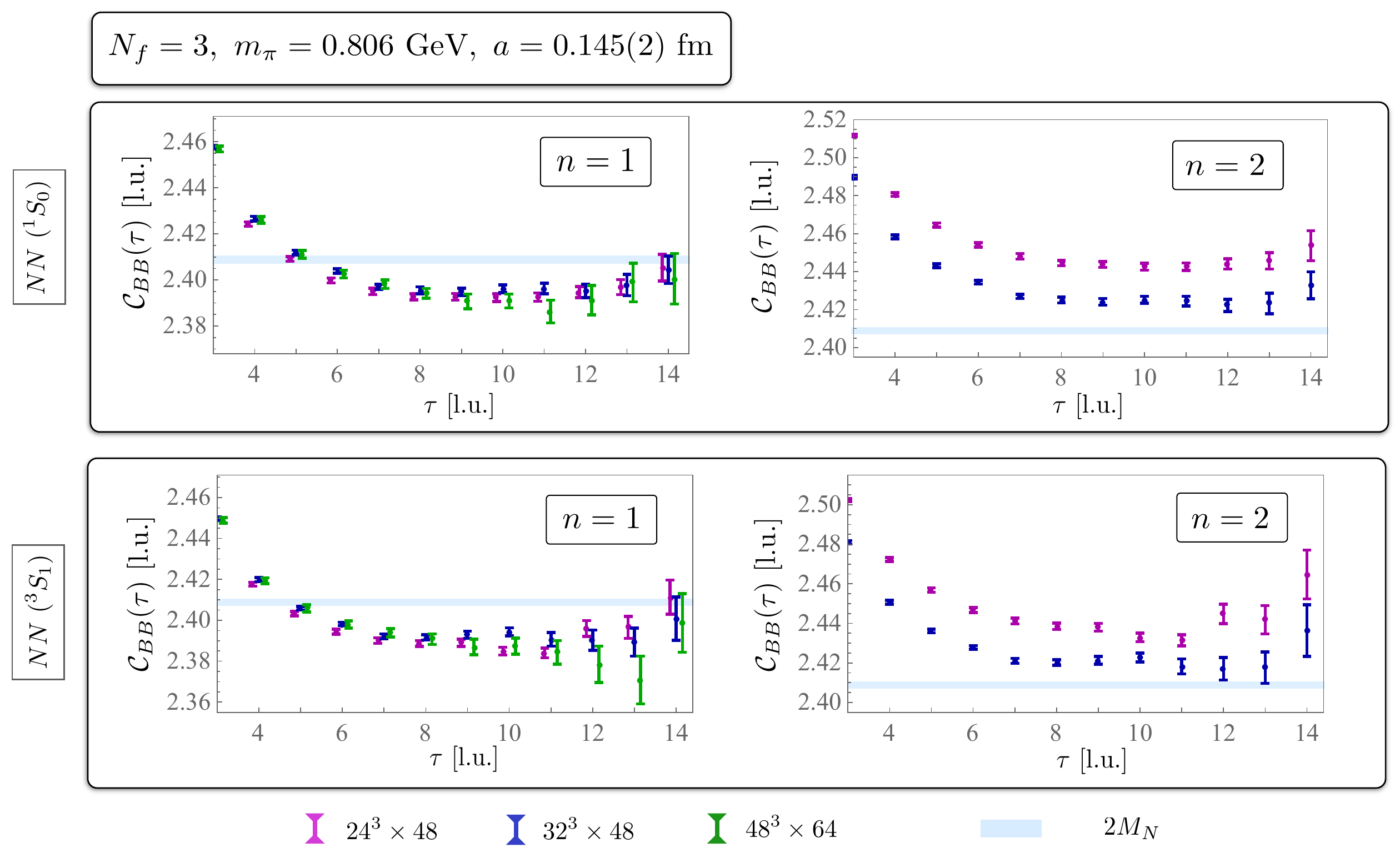}
\caption{A comparison of the effective-mass plots of two-nucleon channels at rest, for the lowest-lying states ($n=1$) in three different lattice volumes in the left panels, and for the second lowest-lying states ($n=2$) in two different lattice volumes in the right panels, presented in Ref.~\cite{Wagman:2017tmp}. The points from different volumes in the panels on the left have been slightly shifted in the time direction for display purposes. The negligible volume dependence of the ground-state EMPs compared with those for the second states in both channels is a signature of a bound state whose size is small compared with the spatial extent of the volume. Quantities are expressed in lattice units (l.u.). Permission to use the figure is granted by the NPLQCD collaboration.}
\label{fig:nn-vol-dep}
\end{figure}

A further argument presented by the NPLQCD collaboration in Refs.~\cite{Wagman:2017tmp, Beane:2017edf} that refutes the claim of the existence of ``fake plateau'' in their calculations concerns the volume-dependence of the energies. In particular, Refs.~\cite{Wagman:2017tmp, Beane:2017edf} present the example of the two-nucleon correlation functions in the spin singlet, $^1S_0$, and spin-triplet, $^3S_1$, channels, from calculations with three degenerate quark flavors, $N_f=3$, quark masses corresponding to $m_{\pi} \approx 806~\tt{MeV}$, a single lattice spacing $a = 0.145(2)~\tt{fm}$, and three different volumes $L \approx 3.4, 4.5$ and $6.7~{\tt{fm}}$, where $L$ denotes the spatial extent of the lattice. This is an important case as there exist discrepancies between the results obtained by the HAL~QCD collaboration using the potential method and those by the NPLQCD collaboration based on the method of ground-state saturation. While the NPLQCD collaboration (as well as the CalLatt collaboration~\cite{Berkowitz:2015eaa}) reports on the existence of bound states in these channels, the HAL~QCD collaboration finds no bound states. HAL~QCD's explanation for such discrepancy is the occurrence ``fake plateaus'' in these channels. However, according to the NPLQCD collaboration, a fake single-exponential behavior (at the times they consider a ground-state saturation to have occurred) can not be the case, as the correlation functions at volumes that differ by at most a factor $8$ coincide, signaling an exponentially small volume dependence that is a feature of a bound state enclosed in a finite volume, see Fig.~\ref{fig:nn-vol-dep}. If the presence of a constant behavior in the EMPs at these moderately early times is the result of peculiar cancelations among multiple exponential terms with nearly equal energies, it is extremely unlikely that for the same cancellations  to occur at different volumes, and with the use of different sink operators, at these times. For comparison, the second energy levels extracted in this study show a large volume dependence, consistent with the power-law volume dependence of the states above the threshold for the ground-state energy of two noninteracting nucleons.

}

\item{A second concern expressed by the HAL~QCD collaboration is the inconsistency between the value of the ground-state energies extracted from the plateau region of the EMPs corresponding to the quantity
\begin{eqnarray}
R(\tau;\mathbf{d})\ = \ \frac{C_{\hat{\mathcal{O}}_{NN},\hat{\mathcal{O}}'_{NN}}(\tau;\mathbf{d})}{\left[C_{\hat{\mathcal{O}}_N,\hat{\mathcal{O}}_N'}(\tau;\mathbf{0})\right]^2} 
=
\mathcal{A}_1e^{-(E^{(0)}_{NN}-2M_N)\tau}
\times \frac{ 1+\mathcal{A}_2e^{-(E^{(1)}_{NN}-E^{(0)}_{NN})\tau}+\dots}{ \left[1+\mathcal{A}_3e^{-(E^{(1)}_{N}-M_N)\tau}+\dots \right]^2},
\label{eq:R}
\end{eqnarray}
for two difference sink operators used: wall sources and (exponentially) smeared sources. Here, $\hat{\mathcal{O}}_N$ and $\hat{\mathcal{O}}_N'$ ($\hat{\mathcal{O}}_{NN}$ and $\hat{\mathcal{O}}_{NN}'$) are interpolating operators for the single(two)-nucleon system and $\mathcal{A}_1$, $\mathcal{A}_2$ and $\mathcal{A}_3$ are known ratios of overlap factors of given states. At late times, a fit to a single exponential can be performed to obtain the energy shift $\overline{\Delta E} \equiv E^{(0)}_{NN}-2M_N$.
\begin{figure}[t]
\centering
\includegraphics[scale=0.4725]{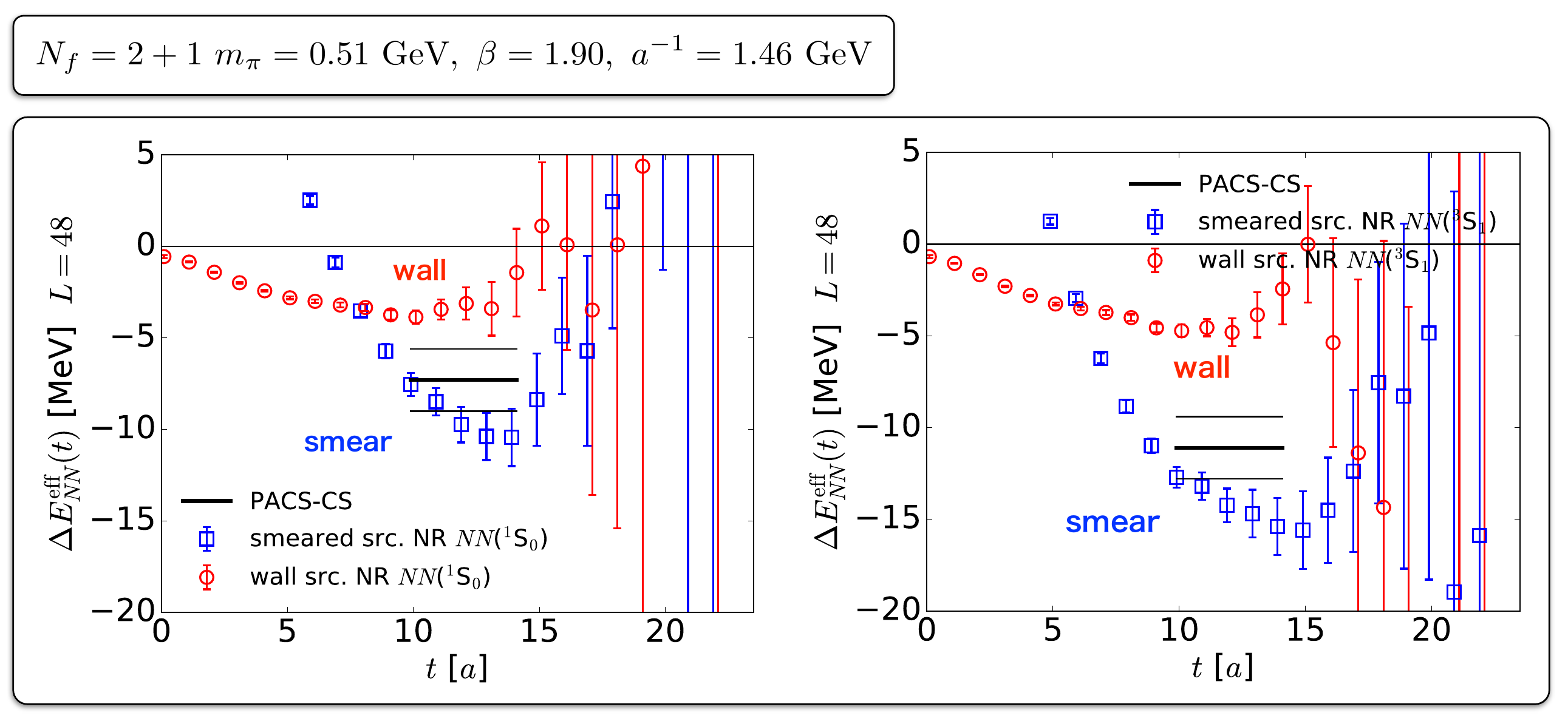}
\caption{The effective energy shifts in the $^1S_0$ (left) and the $^3S_1$ two-nucleon channels (right) obtained in Ref.~\cite{Iritani:2016jie} for both smeared and wall sources, along with plateaus by the PACS-CS collaboration in Ref.~\cite{Yamazaki:2012hi}, showing a (naive) discrepancy in the value of the difference in the ground-state energy of two interacting nucleons and that of two free nucleons. Permission to use the figure is granted by S. Aoki.}
\label{fig:hal-smear-wall}
\end{figure}
An example of such discrepancy is shown in Fig.~\ref{fig:hal-smear-wall}, where a naive assignment of the earliest constant region in the EMPs, according to Ref.~\cite{Iritani:2016jie}, does not agree between the smeared and wall sources. Shown in the diagrams are also the results reported by the PACS-CS collaboration for the ground-state energy of these systems on the same gauge-field configurations~\cite{Yamazaki:2012hi}. HAL~QCD finds this feature to be a signal for the appearance of ``fake plateaus'' in the PACS-CS study and concludes that the true plateaus would form only much later, when the EMP is completely dominated by the noise. They conclude that this result likely does not correspond to a bound state, a result that has been already reached by the HAL~QCD collaboration in these channels at similar values of the quark masses using the potential method.

\begin{figure}[t]
\centering
\includegraphics[scale=0.505]{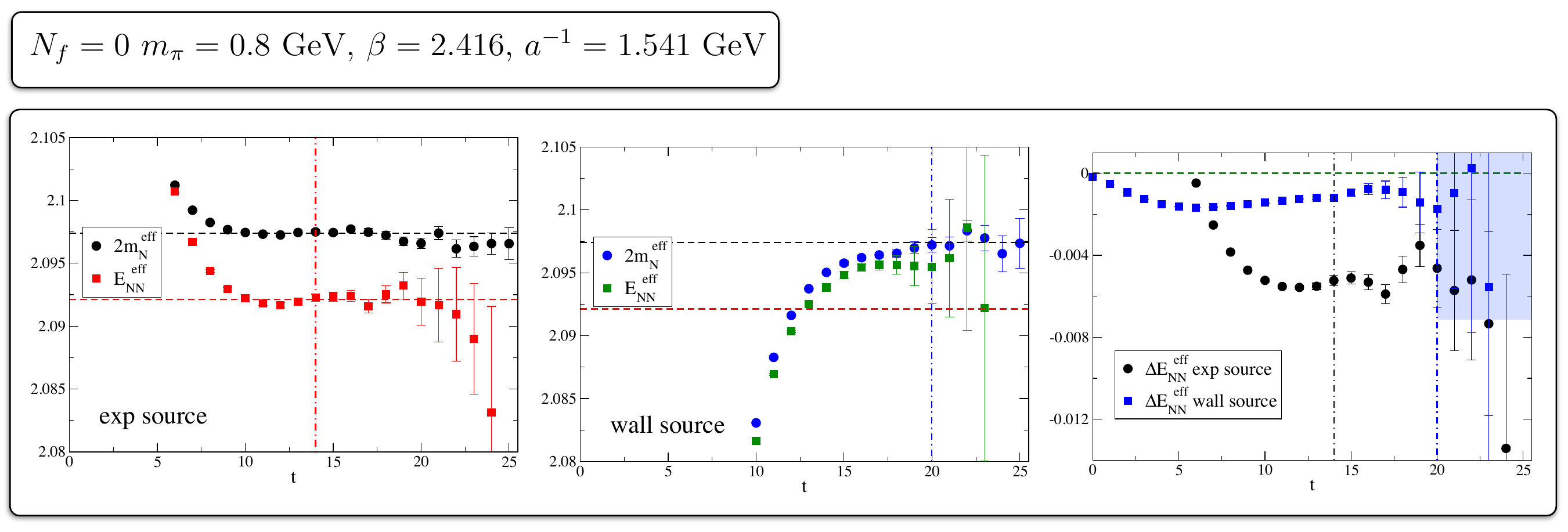}
\caption{The comparison between the EMPs of the nucleon ($\times 2$) and a two-nucleon system using exponentially smeared sources (left panel) and the wall sources (middle panel). These are combined in a single plot for the difference in the ground-state energy of two interacting nucleons and that of two free nucleons (right panel). The dashed lines represent the earliest times a constant value is reached in the EMPs. Given that the wall source plateaus to the ground state much later than the smeared source, there is no discrepancy between the ground-state energy shifts extracted from the smeared source (whose corresponding fit value is not shown) and the wall source (whose corresponding value and uncertainty is roughly estimated to be the blue band). Permission to use the figure is granted by T. Yamazaki.}
\label{fig:pacs-smear-wall}
\end{figure}

A thorough investigation of this feature was conducted by the PACS-CS collaboration in the past year, the results of which were reported in this conference~\cite{Yamazaki:2017jfh} (see also~\cite{Yamazaki:2017euu}). This high-statistics study with $N_f=0$ and a pion mass of $0.8~\tt{MeV}$, concluded that  the ground-state energy shifts, $\overline{\Delta E}$, obtained from the EMPs that used smeared sources are reliable despite the earlier times used in fitting the observed plateaus. On the other hand, it is argued that in the case of wall sources, one cannot associate the observed plateaus in the energy-difference plots with $\overline{\Delta E}$ at early times. The reason is that unless both the numerator and the denominator in Eq. (\ref{eq:R}) are single exponential forms at a given time (corresponding to respective ground states of the two-nucleon and single-nucleon systems), the single-exponential behavior (constant behavior in the corresponding EMPs) can not be associated with the energy difference between the ground states of two interacting nucleons and that of the two free nucleons, $\overline{\Delta E}$. As is shown in Fig.~\ref{fig:hal-smear-wall}, for example, the earliest time at which the single and two-nucleon EMPs with smeared sources can be assured to be in their respective ground states is $t=14$ (in lattice units). As a result, the observed constant behavior in the EMP corresponding to the energy difference at $t \geq 14$ can be identified as $\overline{\Delta E}$. On the other hand, for the single and two-nucleon EMPs with wall sources, ground-state saturation is reached at much later times, $t \geq 20$. The EMP corresponding to the energy difference appears to present a plateau at earlier times, corresponding to a peculiar cancellations among the numerator and the denominator in quantity $R(\tau;\mathbf{d})$. As a result, the observed discrepancy between the plateaus associated with the smeared and wall sources should not be interpreted as a failure of the method based on the ground-state saturation, but instead the poor overlap of the wall sources used to interpolate ground states of the nucleon and two-nucleon systems, see also Ref.~\cite{Savage:2016egr}. In fact, once the constraint $t \geq 20$ is taken into account in fitting the constant region in the EMPs of the wall sources, the obtained value for the energy difference agrees with that obtained with smeared sources, although with a much larger uncertainty (the blue band in the lower-left panel of Fig.~\ref{fig:pacs-smear-wall}\footnote{Addition by the author.}).
\begin{figure}[t]
\centering
\includegraphics[scale=0.480]{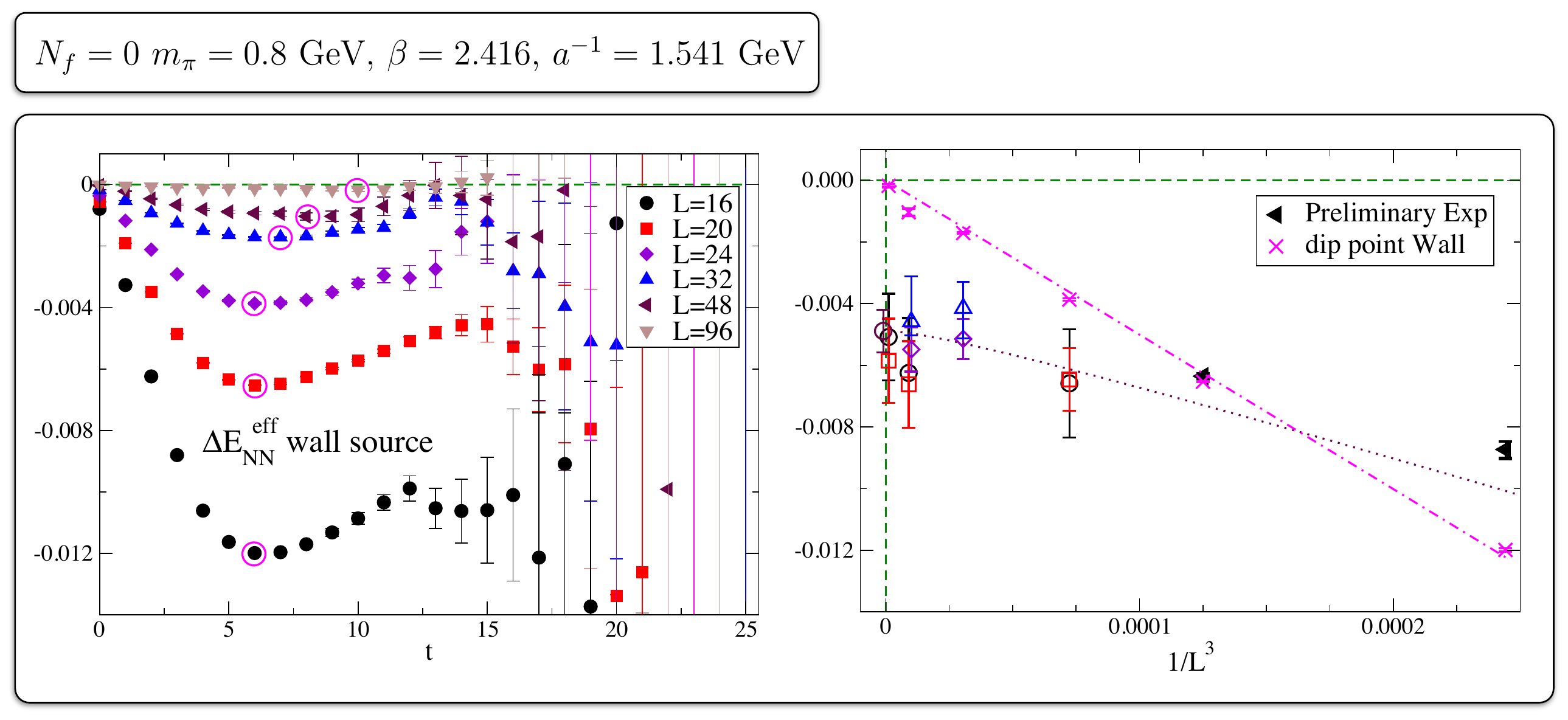}
\caption{The effective energy shifts of a two-nucleon system using the wall source on several volumes are shown in the left, along with the volume dependence of the energy shifts in the right, as presented in Ref.~\cite{Yamazaki:2017euu}. The dash-dot line is to guide the eye. While the ``energy shifts'' corresponding to the dip points of the wall source follow a power-law dependence in volume, the true energy shifts obtained from the ground states of the exponentially smeared sources have a suppressed volume dependence within uncertainties and approach a constant negative value in the infinite-volume limit.  Permission to use the figure is granted by T. Yamazaki.}
\label{fig:pacs-vol}
\end{figure}

Another worthwhile observation made by the PACS-CS collaboration further demonstrates the problematic feature of the wall sources, and enhances confidence in the ground-state energies of two-nucleon systems obtained from the smeared sources. As is shown in Fig.~\ref{fig:pacs-vol}, if one naively assumes a ground-state saturation at the dip points of the EMPs with the wall sources, the corresponding energy shifts will follow a power-law volume dependence. As the ground-state saturation occurs much later in these plots (compare with Fig.~\ref{fig:pacs-smear-wall}), such power-law volume dependence is only a signature of admixture with the excited states. This is in contrary to the results obtained from the smeared sources (that plateau to the ground state much earlier), which present a negligible volume dependence. Additionally, these energy shifts approach a negative value in the infinite-volume limit, a feature that is expected for a bound state, and is consistent with the argument presented by the NPLQCD collaboration, see Fig.~\ref{fig:nn-vol-dep}. This observation strongly suggests that those calculations that have relied primarily on wall nuclear sources in obtaining their results, such as those in Refs.~\cite{Ishii:2006ec, Aoki:2008hh, Nemura:2008sp, Aoki:2009ji, Inoue:2011ai, HALQCD:2012aa, Yamada:2015cra, Miyamoto:2017tjs, Gongyo:2017fjb}, must be re-examined for possible excited-state contaminations at early times, contaminations which can lead to wrong conclusions regarding the presence or absence of a bound state in these channels.}

\item{Finally, the HAL~QCD collaboration has further proposed several ``sanity'' checks~\cite{Iritani:2017rlk} on the two-baryon scattering amplitudes that are obtained with the use of L\"uscher's finite-volume formula, and has concluded that all studies to date that have relied on a ground-state saturation fail to pass these checks, hence signaling the occurrence of ``fake plateaus'' in the two-baryon systems. One sets of results that do not pass the checks, \emph{according to the HAL~QCD's investigation}, is that by the NPLQCD collaboration in Ref.~\cite{Beane:2013br} on the two-nucleon systems at a $SU(3)$-symmetric point with a pion mass of $m_{\pi} \approx 806~\tt{MeV}$. Within the last year, NPLQCD collaboration has conducted a thorough examination~\cite{Beane:2017edf} of these results and new results obtained for other two-baryon channels at this pion mass~\cite{Wagman:2017tmp}, and has established their validity with regards to the checks proposed in Ref.~\cite{Iritani:2017rlk}, a conclusion that disagrees with that reached by the HAL~QCD collaboration. Consequently, the conclusions presented in Ref.~\cite{Iritani:2017rlk} concerning other studies must be fully examined before a definite statement can be made regarding the state of the results in literature for multi-nucleon systems. Nonetheless, these checks (some to be taken with more caution) are quite useful in establishing either the validity of LQCD determination of the finite-volume spectra or the assumptions made about the low-energy parametrization of the scattering amplitude in a given hadronic channel and  given the values of the quark masses of the calculation.
}
\end{itemize}

In summary, it appears that the concerns expressed by the HAL~QCD collaboration do not pose issues on most of the existing multi-nucleon calculations that are based on the assumption of a ground-state saturation. Given the energy gaps existing in the lowest-lying spectra of these systems, the use of optimal source and sink operators and the caution given to fitting the ratio of correlation functions, the lowest-lying spectra of multi-nucleon systems reported in most studies appear to be robust. Checks on scattering amplitudes obtained from these spectra are worthwhile but various group must yet establish whether the conclusions in Ref.~\cite{Iritani:2017rlk} concern their work.

\section{Baryon-baryon interactions}\label{sec:BB}
\noindent
The interactions among hyperons (counterparts of nucleons with one or more valence strange quarks), are crucial in understanding the role of these baryons in the dense matter. Given the short lifetime of hyperons and hypernuclei, it is extremely challenging to push the precision of phenomenological constraints on their interactions to a level comparable to that of nucleons and nuclei. While there is a  dedicated experimental program to physics of strangeness, it is conceivable that LQCD can make significant contribution in this area in the upcoming years. Given the larger mass of the strange quark, LQCD computations involving hyperons are less costly than those involving only the nucleons. With the expected impact on phenomenology, several groups have invested in obtaining the interactions among baryons in recent years form LQCD. Two studies were reported in this conference which will be briefly reviewed in this section.

The HAL~QCD collaboration is conducting a LQCD study of systems involving two octet baryons, as well as those involving the maximally-strange $\Omega$ baryon, at nearly physical quark masses, corresponding to a pion mass of $m_{\pi} = 146~\tt{MeV}$~\cite{Gongyo:2017fjb}. Their method consists of finding the equal-time Nambu-Bethe-Salpeter wave function $\psi(\bm{r})$ from the ratios of the following LQCD correlation functions,
\begin{eqnarray}
R(\bm{r},t)=\frac{\braket{0\left | B(\bm{r},t)B(\bm{0},t)\mathcal{J}^{\dagger}_{BB}(0)  \right |0}}{\left(\braket{0\left | \sum_{\bm{r}}B(\bm{r},t)\mathcal{J}^{\dagger}_{B}(0)  \right |0}\right)^2}=\sum_{n} Z_n\psi(\bm{r})e^{-\Delta E_n t}+\mathcal{O}\left( e^{-\delta_B t} \right),
\label{eq:NBS}
\end{eqnarray}
and consequently a non-local two-baryon potential $U(\bm{r},\bm{r}')$,
\begin{eqnarray}
\left [ \frac{\nabla^2}{M_B}-\frac{\partial}{\partial t}+\frac{1}{4M_B}\frac{\partial^2}{\partial t^2} \right ]R(\bm{r},t)=\int d\bm{r}' U(\bm{r},\bm{r}') R(\bm{r}',t),
\label{eq:potential}
\end{eqnarray}
which can then be expanded in a low-order velocity expansion with local coefficients, $U(\bm{r},\bm{r}')=V(r)\delta(\bm{r},\bm{r}')+\dots$. The obtained approximation to the potential can be used to solve the quantum-mechanical scattering problem in a finite volume. Here, $B(\bm{r},t)$ is the sink operator for a baryon and $\mathcal{J}^{\dagger}_{BB}$ ($\mathcal{J}^{\dagger}_{B}$) is the source operator for two baryons (a single baryon). $Z_n$ is the ratio of the overlap factors for the $n^{th}$ two-baryon state and the ground single-baryon state. $\Delta E_n$ is the difference between the $n^{th}$ energy eigenvalue of two baryons in a finite volume and twice the mass of the baryon, $M_B$. Finally, $\delta$ denotes the first excitation gap of the single baryon.
\begin{figure}[t]
\centering
\includegraphics[scale=0.355]{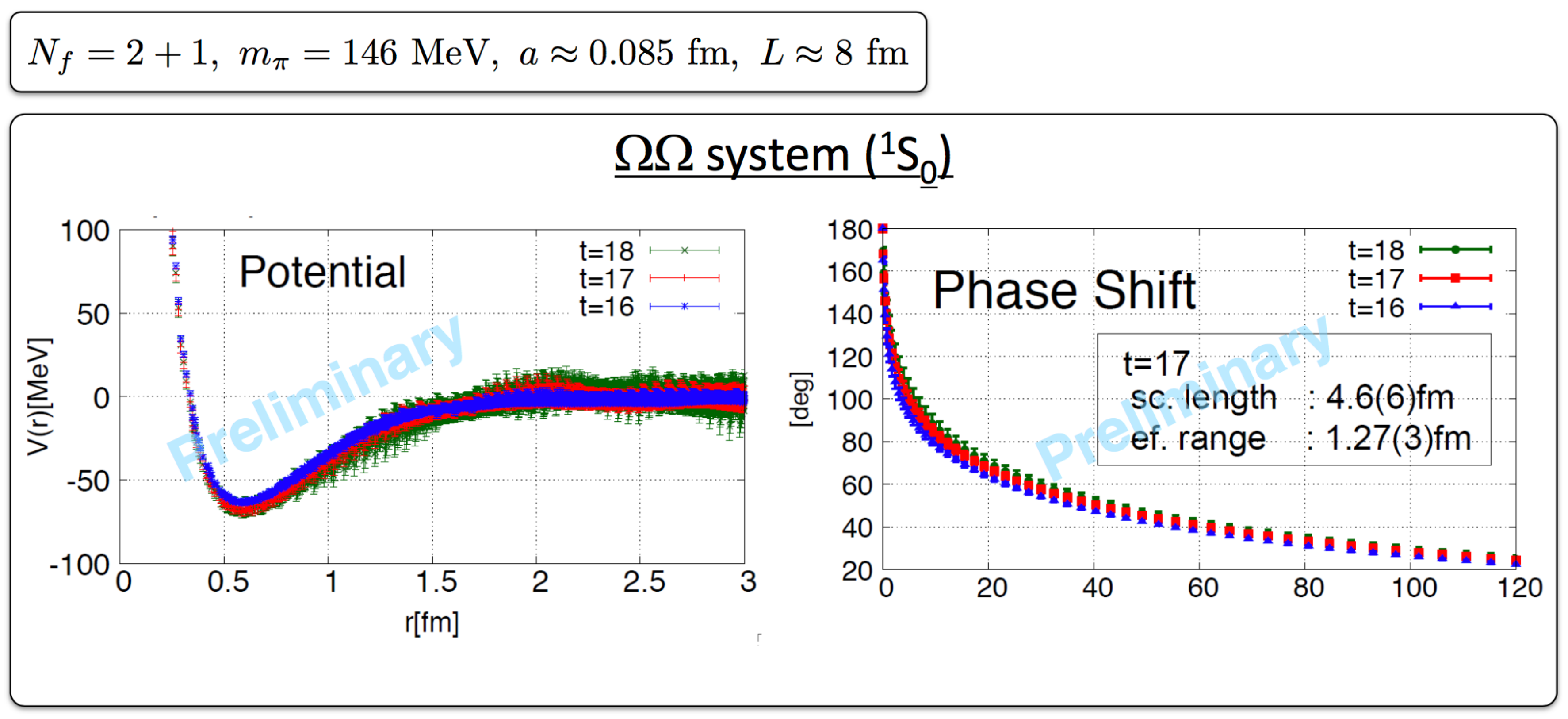}
\caption{Preliminary results by the HAL~QCD collaboration for the potential as a function of source-sink separation (left), and the extracted s-wave scattering phase shift as a function of the center-of-mass (CM) energy (right) in the $^1S_0$ $\Omega\Omega$ channel. For updated results see Ref.~\cite{Gongyo:2017fjb}. Permission to use the figure is granted by T. Doi.}
\label{fig:bb1-hal}
\end{figure}

The results obtained with this method are guaranteed to produce the infinite-volume scattering amplitude at the corresponding energy eigenvalues of the two-baryon system, by construction. However, as both the wavefunction and the potential are not physical observables, and in particular given their operator dependence at short distances (see Eq. (\ref{eq:NBS})), the potential is in general energy dependent, and the connection to scattering amplitudes is obscure away from the finite-volume energy eigenvalues \cite{Beane:2010em, Yamazaki:2017gjl}. While the convergence of the velocity expansion is generally checked by the HAL~QCD collaboration~\cite{Iritani:2017wvu}, their recent study of the $\rho$ resonance using the potential method~\cite{Kawai} puts serious doubts on the assumption of negligible operator dependence in this construction, a feature that demands further investigations. Finally, the potential method, according to the HAL~QCD collaboration, does not require a ground-state saturation in the two-hadron sector, as once the single hadron is in its ground state, a single potential can be obtained for the tower of two-hadron states below the inelastic thresholds. This, once again, assumes an energy-independent potential which generally is not a feature of the potential obtained from LQCD correlation functions with an arbitrary sink structure for the two hadrons, and the distance at which the energy dependence is insignificant is not known a priori for calculations with a given quark mass. Nonetheless, assuming that such systematic uncertainties are under control, one may proceed to constrain the (model-dependent) potentials and the scattering phase shifts determined from these potentials in various two-baryon channels.
\begin{figure}[t]
\centering
\includegraphics[scale=0.485]{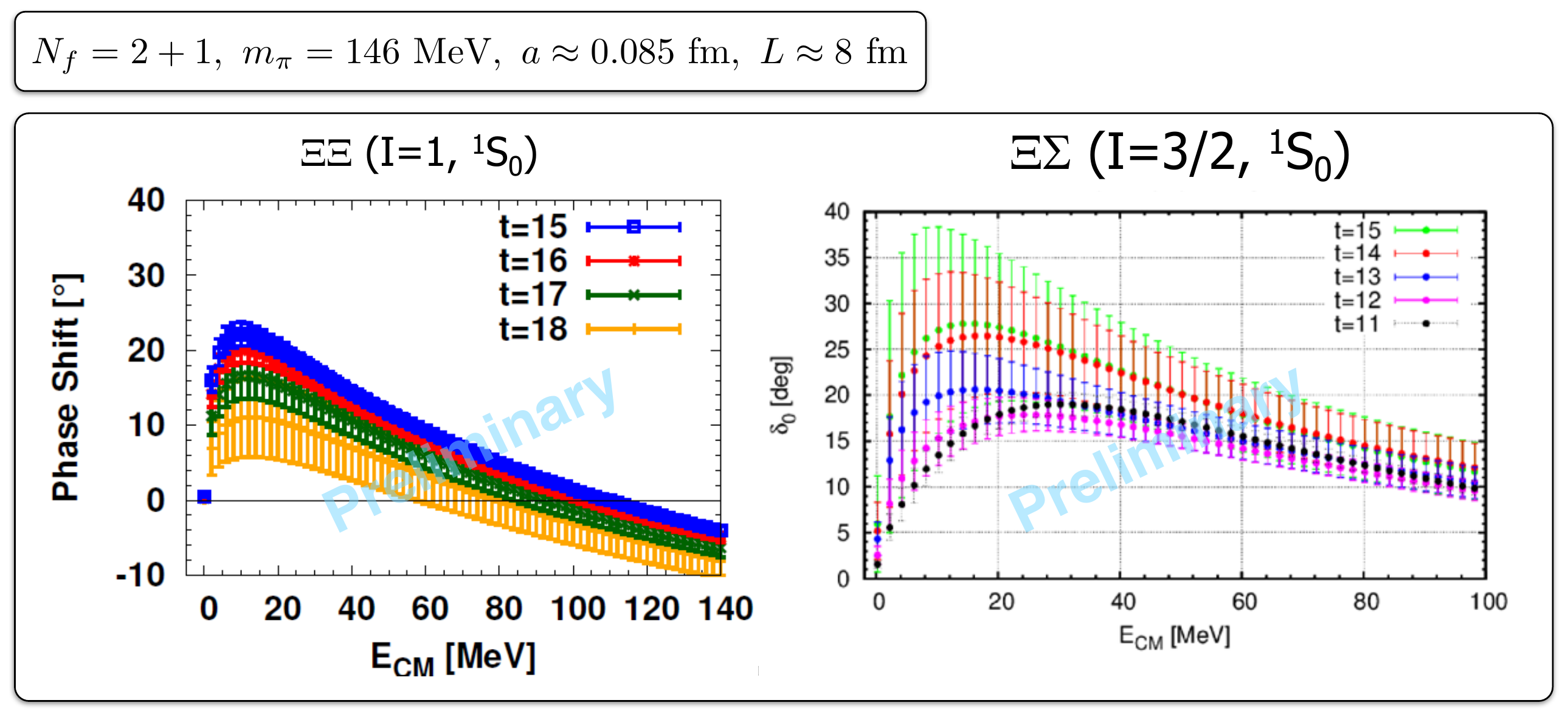}
\caption{Preliminary results by the HAL~QCD collaboration for the s-wave scattering phase shift as a function of the CM energy in the $^1S_0$ $\Xi\Xi$ channel (left) and the $^1S_0$ $\Sigma\Xi$ channel (right), obtained using the potential method. Permission to use the figure is granted by T. Doi.}
\label{fig:bb2-hal}
\end{figure}
\begin{figure}[t]
\centering
\includegraphics[scale=0.41]{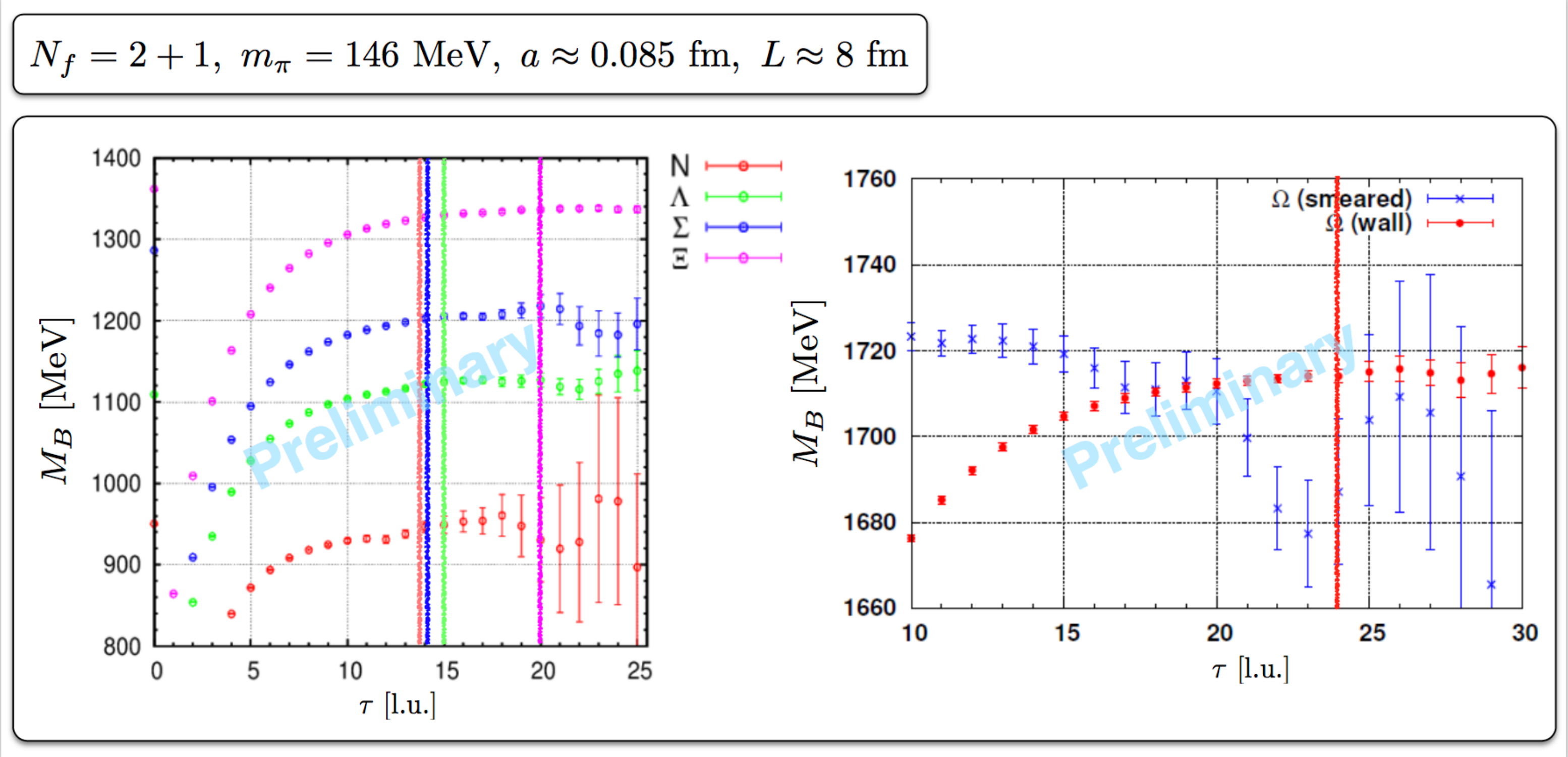}
\caption{EMPs of the $N$, $\Lambda$, $\Sigma$ and $\Xi$ baryons in left and that of the $\Omega$ baryon, obtained by the HAL~QCD collaboration. The vertical lines mark the author's estimate of the start of the plateau region in EMPs. Permission to use the figure is granted by T. Doi.}
\label{fig:b-hal}
\end{figure}

Among the preliminary results that have been obtained by the HAL~QCD collaboration from their 2+1-flavor LQCD calculations with a spatial volume of $(8.1~\tt{fm})^3$ at a lattice spacing of $0.0846~\tt{fm}$, are the phase shifts in: 1) the $^1S_0$ and $^5S_2$ $\Omega \Omega$ channels, both pointing to a strong attraction in these channels (see Fig.~\ref{fig:bb1-hal}) with the $^1S_0$ channel being near unitarity~\cite{Gongyo:2017fjb}, 2) the $^1S_0$ $\Xi \Xi~(I=1)$ and $^1S_0$ $\Xi \Sigma~(I=\frac{3}{2})$ channels, both showing a strong attraction (see Fig.~\ref{fig:bb2-hal}) but possibly support no bound state, and 3) the $^3S_1-{^3}D_1$ $N \Sigma$ channel for which the statistical uncertainties are still significant to allow any conclusion to be made. One has to keep in mind that the ground-state saturation in the single-baryon correlation function must still be assured before any two-particle potential can be extracted. With the EMPs of the octet baryons and that of $\Omega$ in Fig.~\ref{fig:b-hal}, it is clear that all the existing results must be updated such that the extraction of the potential at much later times than shown is possible or else the results are severely contaminated by the excited states of the baryon. For example, with the (author's) rough estimate of the start of the plateau region in the baryon EMPs presented by the HAL~QCD collaboration, the $\Xi$ and $\Omega$ baryons are in their ground state at $t \geq 20$ and $t \geq 23$ in lattice units, at which the potentials are yet to be obtained. 

\begin{figure}[t]
\centering
\includegraphics[scale=0.525]{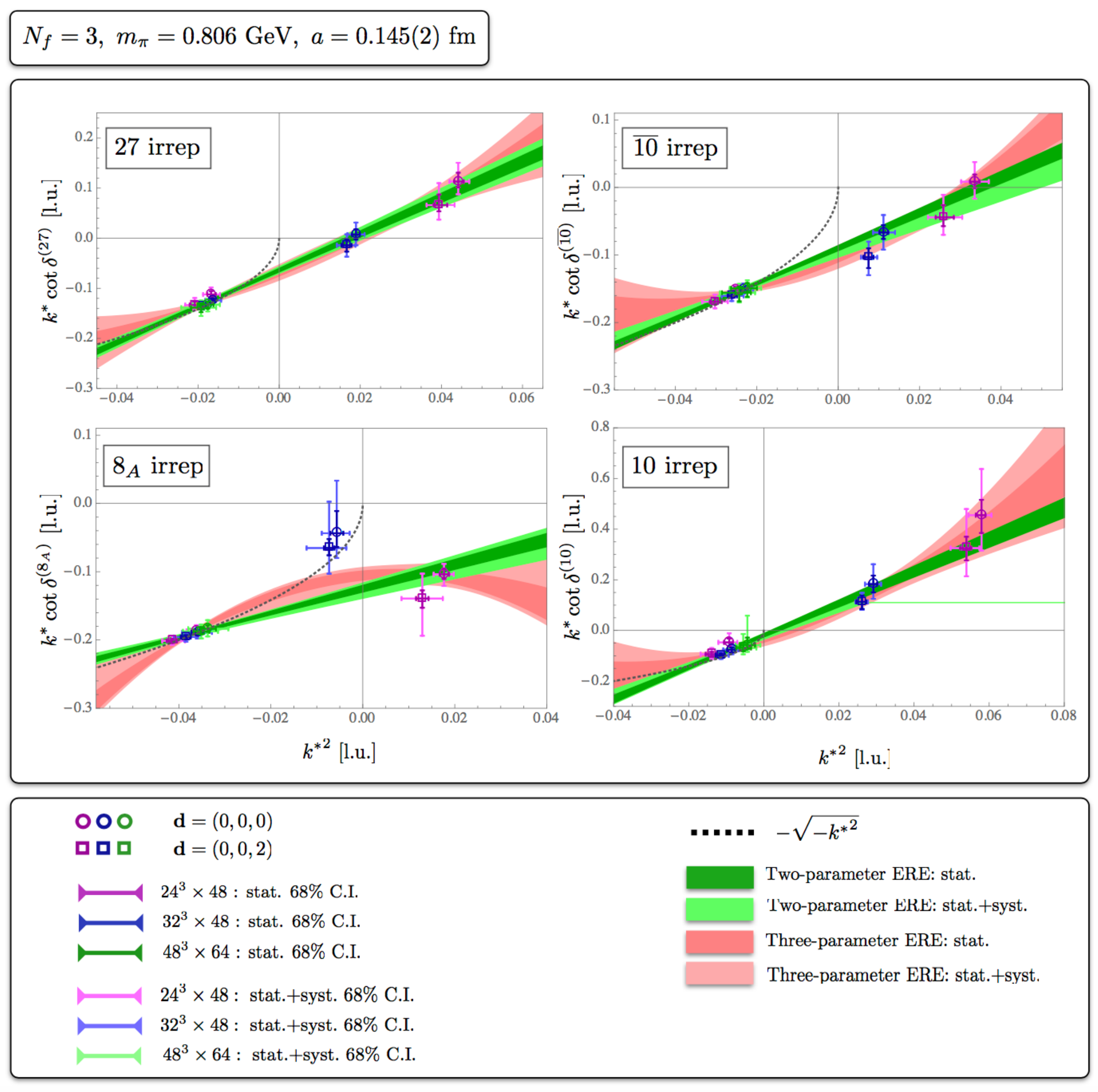}
\caption{$k^*\cot \delta$ versus the square of the CM momentum of the two baryons, ${k^*}^2$, in four  irrep of $SU(3)$~\cite{Wagman:2017tmp}. The bands represent fits to the two and three-parameter effective range expansions. Quantities are expressed in lattice units (l.u.). Permission to use the figure is granted by the NPLQCD collaboration.}
\label{fig:bb-nplqcd}
\end{figure}

The NPLQCD collaboration has also performed a study of two octet baryons with three degenerate quark flavors, $N_f=3$, quark masses corresponding to $m_{\pi} \approx 806~\tt{MeV}$, a single lattice spacing, $a = 0.145(2)~\tt{fm}$, and with three different volumes, $L \approx 3.4, 4.5$ and $6.7~{\tt{fm}}$~\cite{Wagman:2017tmp}. The goal of this study was to extend the previous work by the collaboration on the two-nucleon interactions~\cite{Beane:2012vq, Beane:2013br} to other two-baryon channels. The merit of such study at a large value of quark masses is that high precision results are possible, allowing conclusive statements to be made about the physics in play when the input parameters of the SM are varied, and setting the stage for calculations at lower quark masses. In this study, the scattering amplitudes at low energies are obtained by L\"uscher's method~\cite{Luscher:1986pf, Luscher:1990ux}. When the $S$-wave scattering is dominant compared with higher partial waves, the $S$-wave scattering phase shift, $\delta_S$, can be accessed from the finite-volume energy eigenvalues through the relation
\begin{eqnarray}
k^*\cot\delta_S=4 \pi
c_{00}^{\mathbf{d}}(k^{*2};  L).
\label{eq:QC}
\end{eqnarray}
Here, $k^*$ is the relative momentum of each baryon in the CM frame and $\mathbf{d}$ denotes the total CM momentum of the system in units of $2\pi/L$. $c_{00}^{\mathbf{d}}(k^{*2};  L)$ is a known kinematic function related to the three-dimensional zeta function. As a consequence of the proximity to the nonrelativistic NR limit, the leading contamination to the $S$-wave L\"uscher's equation for boost vectors with even components arises from nonvanishing $G$-wave interactions~\cite{Briceno:2013lba, Briceno:2013bda}, which are expected to be suppressed relative to the $S$-wave interactions. The values of $k^*\cot\delta_S$ as a function of $k^{*2}$ are shown in Fig.~\ref{fig:bb-nplqcd} for two-baryon channels belonging to  the $27$ (e.g., $NN~({^1}S_0)$), $\overline{10}$ (e.g., $NN~({^3}S_1)$), $10$ (e.g., $\Sigma^+ p~({^3}S_1)$) and $8_A$ (e.g., $\frac{1}{\sqrt{2}}(\Xi^0n+\Xi^-p)~({^3}S_1)$) irreducible representations (irreps) of $SU(3)$, along with the two-parameter and three-parameter effective range expansion ($k^*\cot\delta_S=-\frac{1}{a}+\frac{1}{2}r{k^*}^2+P{k^*}^4+\dots$) fits (the t-channel cut at this pion mass starts at $k^{*2} \approx 0.088$ in lattice units). $a$, $r$ and $P$ are the scattering length, effective range and the leading shape parameter, respectively. The energies that are input into Eq. (\ref{eq:QC}) are obtained from various analysis techniques, see Ref.~\cite{Wagman:2017tmp} for details, and examples of the EMPs are shown in Fig.~\ref{fig:nn-vol-dep}. As argued in Sec.~\ref{sec:robustness}, the volume dependence of the energies provides overwhelming evidence for the existence of a bound state in each of the $27$, $\overline{10}$ and the $8_A$ irreps, while the systems belonging to the $10$ irrep appear to be near threshold, and the binding energies are determined to be
\begin{eqnarray}
\centering
27 \text{ irrep:}&&~~~B=20.6{}_{(-2.4)}^{(+1.8)}{}_{(-1.6)}^{(+2.8)}~{\tt{MeV}},~~~~~~\overline{10} \text{ irrep:}~~~B=27.9{}_{(-2.3)}^{(+3.1)}{}_{(-1.4)}^{(+2.2)}
 ~\tt{MeV},
 \\
\nonumber\\
10 \text{ irrep:}&&~~~B=6.7{}_{(-1.9)}^{(+3.3)}{}_{(-6.2)}^{(+1.8)}
 ~{\tt{MeV}},~~~~~~8_A \text{ irrep:}~~~B=40.7{}_{(-3.2)}^{(+2.1)}{}_{(-1.4)}^{(+2.4)}
 ~\tt{MeV},
\label{eq:binding-phys-10}
\end{eqnarray}
in physical units, agreeing with those obtained previously for these ensembles~\cite{Beane:2013br, Berkowitz:2015eaa}. The first and second uncertainties in each pair are statistical and systematic uncertainties, respectively.

\begin{figure}[t]
\centering
\includegraphics[scale=0.505]{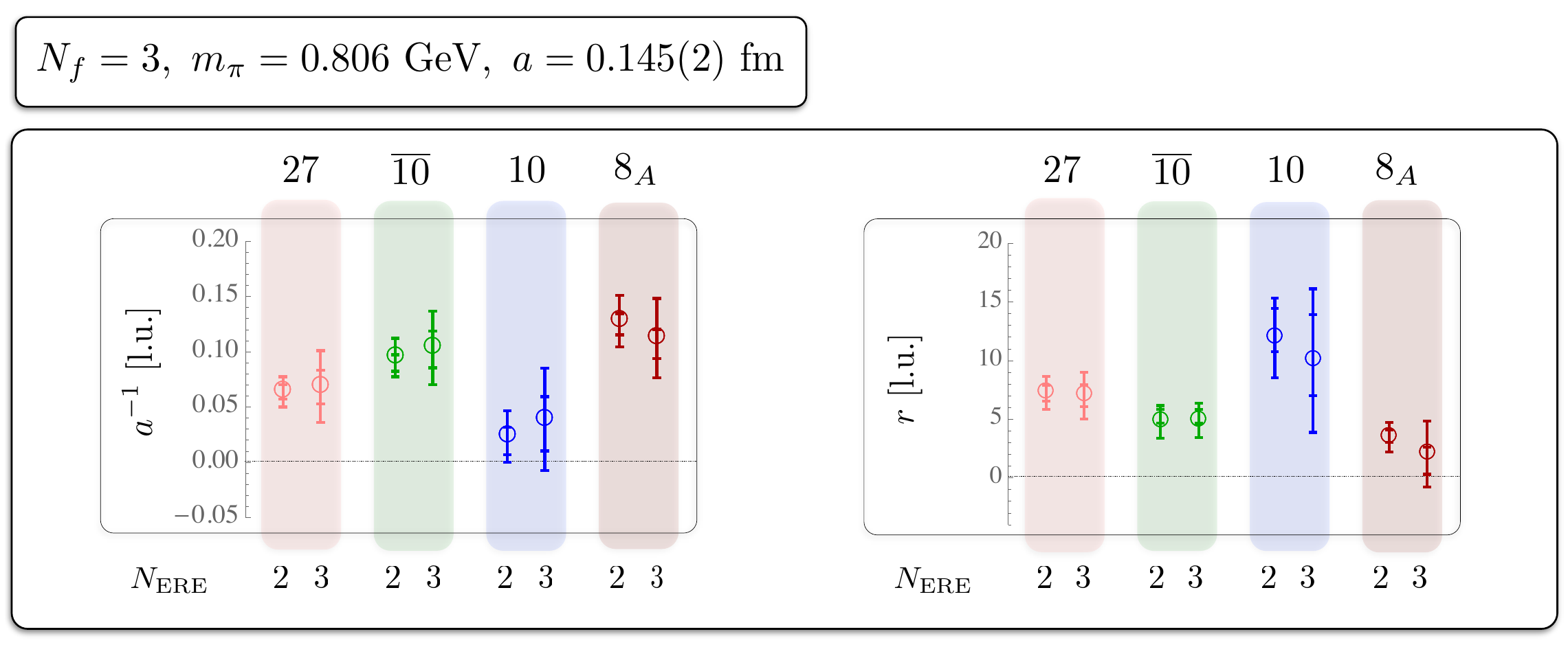}
\caption{A comparison of the values of the inverse scattering length, $a^{-1}$ and effective range, $r$, in the two-baryon channels belonging to the four irreps of $SU(3)$~\cite{Wagman:2017tmp}. These point to a $SU(6)$ spin-flavor symmetry in nuclear and hypernuclear forces. Permission to use the figure is granted by the NPLQCD collaboration.}
\label{fig:ar-nplqcd}
\end{figure}

The values of scattering parameters obtained in these channels are shown in Fig.~\ref{fig:ar-nplqcd}, from which an approximate $SU(6)$ spin-flavor symmetry can be observed, a feature that is predicted in Ref.~\cite{Kaplan:1995yg} to hold for QCD in the limit of a large number of colors, $N_c$, but is obtained directly from QCD for the first time in this study. In particular, the results of Ref.~\cite{Wagman:2017tmp}, supplemented with the assumption of the proximity of the values of scattering parameters in the two remaining irreps of $SU(3)$ to the ones obtained in other channels, enable an extraction of the six $SU(3)$ Savage-Wise coefficients~\cite{Savage:1995kv} of the leading order effective Lagrangian
\begin{eqnarray}
\mathcal{L}_{BB}^{(0)}&=&
-c_1 \text{Tr}(B_i^{\dagger}B_iB_j^{\dagger}B_j)
-c_2 \text{Tr}(B_i^{\dagger}B_jB_j^{\dagger}B_i)
-c_3 \text{Tr}(B_i^{\dagger}B_j^{\dagger}B_iB_j)
\nonumber
\\
&& -c_4 \text{Tr}(B_i^{\dagger}B_j^{\dagger}B_jB_i)
-c_5 \text{Tr}(B_i^{\dagger}B_i)\text{Tr}(B_j^{\dagger}B_j)
-c_6 \text{Tr}(B_i^{\dagger}B_j)\text{Tr}(B_j^{\dagger}B_i),
\label{eq:BB-Lagrangian}
\end{eqnarray}
where $B$ is the well-known octet-baryon matrix and roman indices denote spin indices. Given the somewhat unnaturally large scattering length in the channels studied in this work, the coefficients can be constrained at a proper renormalization scale. The result of this matching is interesting and concludes that only a single $SU(3)$ interaction, the term proportional to $c_5$ in Eq. (\ref{eq:BB-Lagrangian}), is significant. This observation implies an extended symmetry beyond $SU(6)$, and is consistent with the accidental $SU(16)$ symmetry of nuclear and hypernuclear forces predicted in Ref.~\cite{Kaplan:1995yg} in the limit of large $N_c$.

The study in Ref.~\cite{Wagman:2017tmp} is an example of the role of LQCD in constraining effective interactions in hadronic systems. This allows \emph{ab intio} nuclear many-body calculations (those taking nucleons as the degrees of freedom) of larger systems to use effective interactions\footnote{The applicability and convergence of the nuclear EFTs in larger nuclei remain unresolved.} that are tuned with QCD input, thereby extending the impact of LQCD to larger systems than those plausible with current resources. Such tuning of the interactions can be implemented at the level of the (continuum-extrapolated) finite-volume spectra and matrix elements obtained with LQCD and those obtained using \emph{ab intio} nuclear many-body methods in the same volume, and so no conversion of LQCD observables to infinite-volume counterparts will be necessary, see also Ref.~\cite{Beane:2012ey}. An example of such interplay between LQCD, nuclear EFTs and many-body theory is given in Ref.~\cite{Barnea:2013uqa}, successfully demonstrating the roadmap from LQCD to larger nuclei~\cite{Contessi:2017rww}. Here, the leading two and three-body interactions of the pionless EFT~\cite{Chen:1999tn} are constrained using the LQCD results for the binding energies of the deuteron, di-neutron and $^3$He/${^3}$H at a $SU(3)$ flavor-symmetric point with a pion mass of $m_{\pi} \approx 806~\tt{MeV}$, and excluding QED~\cite{Beane:2012vq}, and the tuned interactions are used, using a Green's function quantum Monte Carlo many-body technique~\cite{Barnea:2013uqa}, to obtain the binding energy of $^4$He, for which a LQCD determination exists. The agreement observed serves as a verification and insures that the LQCD binding energies obtained from the assumption of the ground-state saturation in the time intervals used are valid. More importantly, this calculation provides predictions for the $A=5,6$ systems for which no LQCD results were obtained. Recently, such predictions are extended to nuclei as large as $^{16}O$~\cite{Contessi:2017rww}, and high-precision calculations with the inclusion of the next-to-leading order interactions are planned to solidify the conclusions regarding the fate of the nuclear landscape with heavy quarks. This is the framework that is envisaged to be implemented once LQCD inputs near and at the physical values of quark masses are reality, therefore bridging between QCD and the many-body nuclear phenomena.

\section{Nuclear reactions in light nuclear systems}\label{sec:reactions}
\noindent
A major development in the past two years is the calculation of matrix elements of currents with nuclear states. The first nuclear reaction cross section with direct input from LQCD was reported in the last year's conference by the NPLQCD collaboration~\cite{Savage:2016egr}. The reaction studied was a radiative capture process, $np \to d\gamma$, which is dominated by the M1 amplitude in the low-energy regime~\cite{Beane:2015yha}. The implementation of a background magnetic field induced a mixing between the $^3S_1$ and $^1S_0$ two-nucleon states and the resulting low-lying spectrum contained information about the M1 transition amplitude, in particular the solely two-body contribution to the process, characterized by the $l_1$ coupling of the pionless EFT~\cite{Beane:2000fi, Detmold:2004qn}. The obtained constraints on this coupling at two different quark masses allowed an extrapolation to the physical point, with a resulting cross section that fully agreed with experimental value. The collaboration extended this study to the case of axial-vector transitions through which phenomenologically interesting quantities such as the $pp$ fusion cross section and the $\beta\beta$-decay matrix elements could be accessed.  

The more customary implementation of background fields in Ref.~\cite{Beane:2015yha} was improved in Refs.~\cite{Savage:2016kon, Shanahan:2017bgi, Tiburzi:2017iux} (see also Refs.~\cite{Bouchard:2016heu, Berkowitz:2017gql}), to allow for a straightforward isolation of the linear and quadratic responses to the external field at the level of the correlation functions. In this method, first a \emph{compound propagator} is formed,
\begin{eqnarray}
S_{\{\Lambda_1,\ldots \}}(x,y) = S(x,y) + \int dz  S(x,z) \Lambda_1(z)  S(z,y) + \dots,
\label{eq:bfprop}
\end{eqnarray}
where $\Lambda_i$ denotes the background field. Both $\Lambda_i(x)$ and the quark propagator $S(x,y)$ are spacetime-dependent matrices in spinor and flavor space, and $S(x,y)$ is also a matrix in color space. Next, one or more normal propagators is replaced by the compound propagator in the construction of the zero-momentum-projected hadronic correlation function
\begin{eqnarray}
C^{(h)}_{\lambda_u;\lambda_d}(t) 
& = & 
\sum_{\bm x}
\langle 0| \chi_h({\bm x},t) \chi^\dagger_h({\bm 0},0) |0 \rangle_{\lambda_u;\lambda_d}.
\label{eq:bfcorr}
\end{eqnarray}
Here, subscript $\lambda_u;\lambda_d$ denotes the presence of a $u$ and/or $d$ compound propagator with insertions of the background fields with coupling $\lambda_u$ and $\lambda_d$, and superscripts $h$ denotes the quantum numbers of the hadronic interpolating operator, $\chi_h$. This approach is only exact for isovector fields, and only for quantities that are maximally stretched in isospin space.  At the single-insertion level, this corresponds to isovector quantities such as the isovector axial charges of the proton and triton, and the axial matrix element relevant for the $pp\to d e^+\nu_e$ fusion cross section.  With two insertions of the background field, isotensor quantities can be computed exactly. 

\begin{figure}[t]
\centering
\includegraphics[scale=0.475]{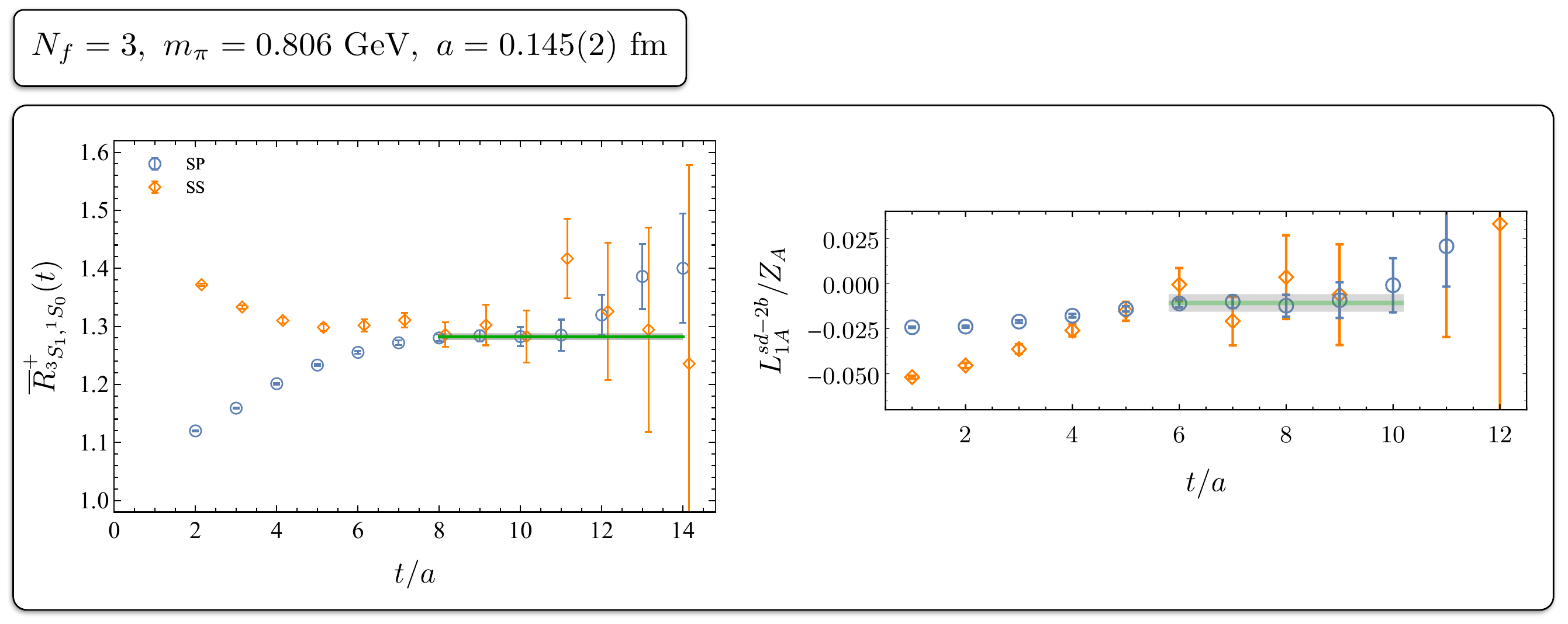}
\caption{The quantity plotted in the left obtains the bare $pp\to d$ transition matrix element in the late-time limit, and the quantity in the right gives the solely two-nucleon contribution to the $pp\to d$ bare transition matrix element at large $t$~\cite{Savage:2016kon}. SP and SS refer to the smeared-point and smeared-smeared source-sink combinations. The horizontal bands show constant fits to the late-time behavior of the SP quantities. The SS points are slightly offset in $t/a$ for clarity. Permission to use the figure is granted by the NPLQCD collaboration.}
\label{fig:ppfusion-nplqcd}
\end{figure}

The study by the NPLQCD collaboration was conducted using the previously generated~\cite{Beane:2012vq} $SU(3)$-symmetric ensembles of gauge-field configurations with quark masses corresponding to $m_{\pi} \approx 806~\tt{MeV}$, a single lattice spacing, $a = 0.145(2)~\tt{fm}$, and a single lattice volume, $L \approx 4.5~{\tt{fm}}$. With the use of an axial-vector background field, besides the axial charge of the proton,\footnote{see Ref.~\cite{Berkowitz:2017gql} for a precise extraction of the axial charge of the proton at the physical point using the same method.} the following results were obtained in two-nucleon systems:

\
\

\noindent
\emph{The $pp$ fusion cross section:} Being the first reaction in the chain of processes that release energy in stars like Sun, $pp$ fusion at low incident velocities is of significant importance in refining solar models. Given its weak strength and the Coulomb barrier between the protons, the rate for this process is currently difficult to measure in experiment. To improve the phenomenological constraints, LQCD can play a role by directly measuring the relevant nuclear matrix element from QCD. Combined with EFTs (enhanced to account for QED)~\cite{Kong:2000px, Butler:2001jj}, such LQCD input can refine the determination of the short-distance two-body axial-current coupling of the pionless EFT, $L_{1,A}$. The calculation performed in the last year with heavy quarks, obtained the matrix element for the axial current in the two-nucleon state with a percent-level uncertainty (see the left panel of Fig.~\ref{fig:ppfusion-nplqcd}), and a rough extrapolation was made to the physical point with the assumption of a mild quark-mass dependence in the solely two-body contribution~\cite{Beane:2015yha}, plotted in the right panel of Fig.~\ref{fig:ppfusion-nplqcd}. The central value and the uncertainties on the quantity $L_{1,A}$ are determined to be: $L_{1,A}= 3.9(0.2)(1.0)(0.4)(0.9)~\tt{fm}^3$ at a renormalization scale $\mu= m_{\pi}$, which is comparable to the current phenomenological value. The uncertainties are statistical, fitting and analysis systematic, quark-mass extrapolation systematic, and a power-counting estimate of higher order corrections in pionless EFT, respectively \cite{Savage:2016kon}. For the calculations towards the physical values of the quark masses, the systematic uncertainty of the present calculation due to the use of extended background fields in time~\cite{Savage:2016kon, Tiburzi:2017iux}, as well as the connection to the infinite-volume matrix elements with an unbound initial state~\cite{Detmold:2004qn, Briceno:2012yi, Briceno:2015tza, Christ:2015pwa} will be addressed.

\
\

\begin{figure}[t]
\centering
\includegraphics[scale=0.35]{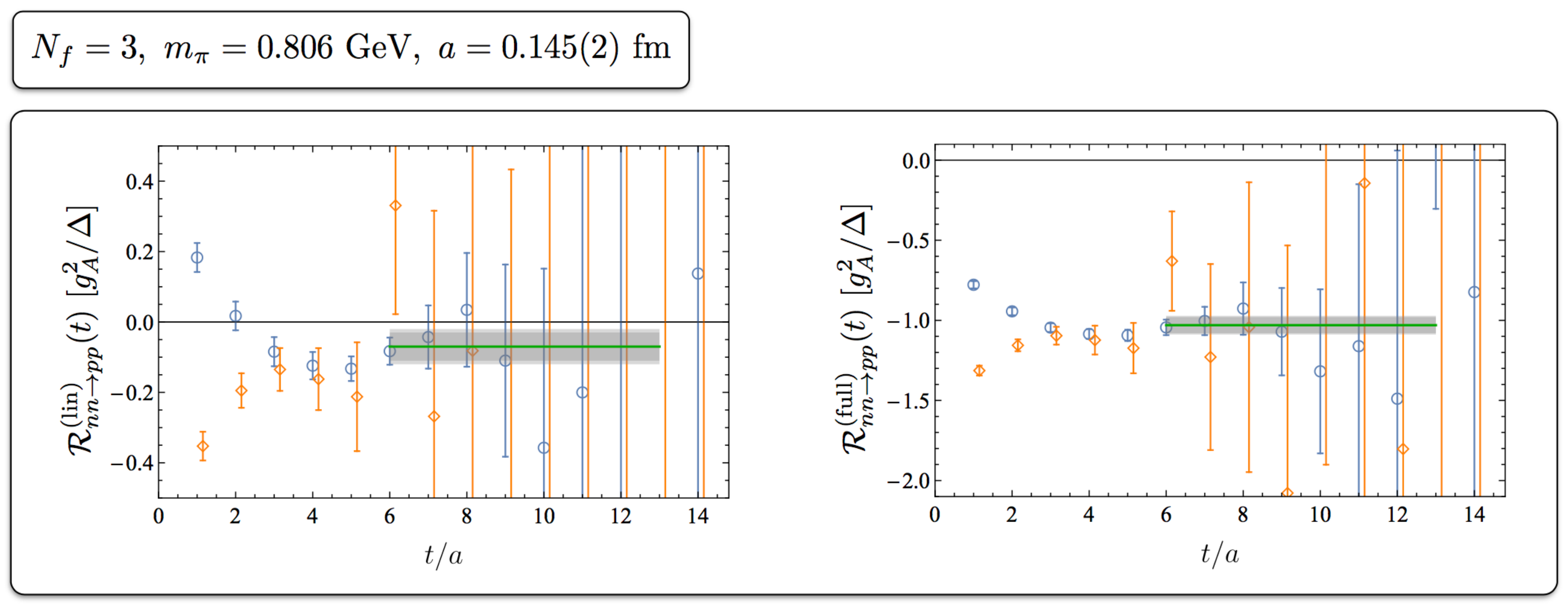}
\caption{The quantity plotted in the left is defined from two-nucleon correlation functions in an axial background field and corresponds to the bare short-distance contribution to the $nn \to pp$ matrix element at late times. The quantity in the right sums the long-distance and short-distance contributions to the matrix element \cite{Shanahan:2017bgi, Tiburzi:2017iux}. SP and SS refer to the smeared-point and smeared-smeared source-sink combinations. The horizontal bands show constant fits to the late-time behavior of the SP quantities. The SS points are slightly offset in $t/a$ for clarity. Permission to use the figure is granted by the NPLQCD collaboration.}
\label{fig:nnpp-nplqcd}
\end{figure}

\noindent
\emph{The matrix element for $nn \to pp$:} In order to complement the searches for the lepton-number violating neutrinoless $\beta\beta$-decay process, it is crucial to constrain, with precision, the relevant nuclear matrix elements. Besides possible new physics contributing to this process at short distances (see Ref.~\cite{Nicholson:2016byl} for LQCD input for this scenario), a well-motivated scenario is the long-distance process due to the bilocal action of the SM left-handed current with the exchange of a light Majorana neutrino. In order to systematically refine the many-body calculations of the corresponding nuclear matrix elements in nuclei, a QCD input is needed for any short-distance (at the scale of the nuclear EFTs) contribution to the process. The simpler process of neutrinofull $\beta\beta$ decay serves as an important benchmark, and the nuclear many-body techniques have concentrated in improving the matrix-element estimates of this process. At the heart of the $\beta\beta$ decay is the inversion of two neutrons to two protons, a process which can not happen in isolation but can be studies in a LQCD calculation of the corresponding QCD matrix element. With two insertions of the axial current, the background-field method of Ref.~\cite{Shanahan:2017bgi, Tiburzi:2017iux} enabled the extraction of this isotensor matrix element at the heavy quark masses of the study (the right panel of Fig.~\ref{fig:nnpp-nplqcd}), along with a constraint on the short-distance non-Born contribution to the matrix element (the left panel of Fig.~\ref{fig:nnpp-nplqcd}). This input was used in a pionless EFT description of the double-weak process $nn \to pp$ to put bounds, for the first time, on a newly identified short-distance two-axial current two-body coupling, $\mathbb{H}_{2,S}= 4.7(2.2)~\tt{fm}$. The contribution arising from this short-distance contribution was found comparable to that from the single-axial current two-body coupling, suggesting that the quenching of $g_A$~\cite{Kumar} in nuclear many-body calculations of the $\beta\beta$-decay rates might not be sufficient for an accurate determination of the matrix elements. Future LQCD studies of the process, with all systematics under control, will provide a crucial input to the neutrinoless $\beta\beta$-decay program.

%

\section{Structure properties of light nuclei\label{sec:structure}}
\noindent
Among the goals of the hadron structure community is to confront the physics output of the planned EIC in the US with theoretical predictions concerning the role of quarks and gluons in the structure of hadrons and nuclei. An important aspect of the program is to establish the significance of medium effects in the single-nucleon properties, and to isolate multi-body effects in the response of the nuclear medium to external currents from the sum of the single-nucleon responses, i.e., the EMC effect~\cite{Aubert:1983xm}. The first LQCD explorations of the EMC effects were conducted last year by the NPLQCD collaboration using axial and gluonic probes, albeit with unphysically heavy quarks, and the following results were reported in this conference:

\
\

\begin{figure}[t]
\centering
\includegraphics[scale=0.34]{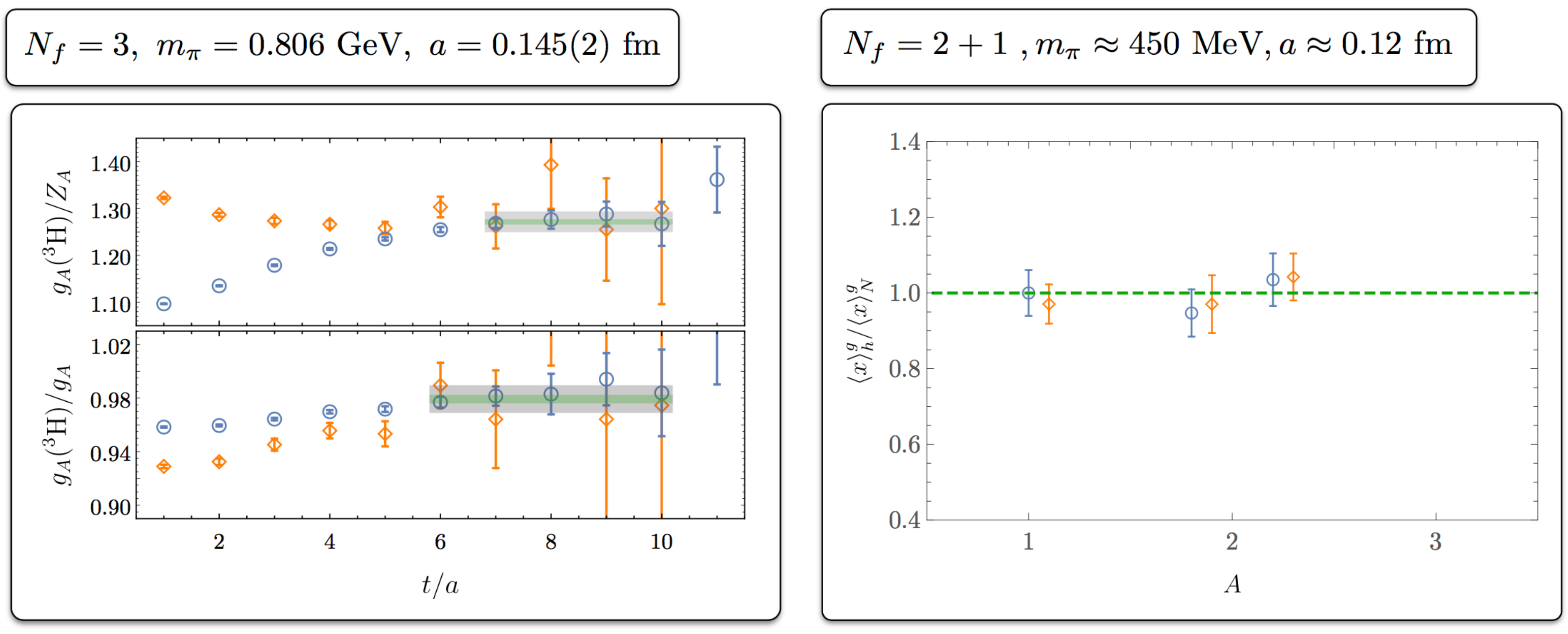}
\caption{The left panel: The ratios of correlation functions that determine the bare isovector axial matrix element in $^3$H (upper panel), and the ratio of the isovector axial matrix element in $^3$H to that in the proton (lower panel) from Ref.~\cite{Savage:2016kon}. SP and SS refer to the smeared-point and smeared-smeared source-sink combinations. The bands correspond to constant fits to the asymptotic behavior. Permission to use the figure is granted by the NPLQCD collaboration. The right panel: Ratios of the gluon momentum fraction in light nuclei, $\braket{x}^g_A$, to that in the proton, $\braket{x}^g$, obtained by the NPLQCD collaboration, allows studies of a potential gluonic EMC effect ~\cite{Winter:2017bfs}. Permission to use the figure is granted by P. Shanahan.}
\label{fig:tritonaxial-nplqcd}
\end{figure}

\noindent
\emph{The axial structure of $^3$He/triton:} The axial charge of $^3$He (equal to that of the triton by isospin symmetry) was obtained by isolating, in the correlation function with compound propagators, the matrix element of the axial current in the ground state of $^3$He. The ratio of the axial charge of $^3$He to that of the proton was found to be $0.979(11)$, exhibiting a nuclear effect in the axial response of this nucleus at the level of $\sim 2$ standard deviations (see the left panel of Fig.~\ref{fig:tritonaxial-nplqcd}). A refined analysis is planned to improve this conclusion regarding the quenching of the axial charge in nuclei. This result subsequently enabled obtaining a constraint on the Gamow-Teller matrix element associated with the tritium $\beta$ decay, a result that at this heavy quark mass was found very close to the physical value, showing a mild quark-mass dependence in this quantity, similar to that found for the axial matrix element in the nucleon.

\
\

\noindent
\emph{Gluonic matrix elements in the nucleon and light nuclei:} Given that gluons carry no electric charge, the experimental investigations of the gluonic distributions inside hadrons and nuclei are significantly challenging, and the upcoming experiments such as the EIC will be devoted to the physics of gluons. Progress in the LQCD study of gluonic observables in hadrons and light nuclei were reported in this conference. In particular, the first results on the unpolarized gluon longitudinal momentum fraction in the nucleon and in light nuclei have emerged at a heavy quark mass corresponding to $m_{\pi}=450~\tt{MeV}$, as shown in the right panel Fig.~\ref{fig:tritonaxial-nplqcd} (see further results in a manuscript that was released subsequently~\cite{Winter:2017bfs}). Such quantity is a signature of the gluonic analog of the EMC effect and is found, from these calculations, to be bounded by $\sim10-20\%$ in light nuclei up to $A=3$. The first moment of the gluonic transversity distribution in the deuteron is a clean measure of gluonic degrees of freedom and is protected from the leading-twist mixing with quark distributions~\cite{Jaffe:1989xy}. This quantity, also studied in spin-1 mesons in Refs.~\cite{Detmold:2016gpy}, is investigated by the NPLQCD collaboration and clear signal was found as heavier pion masses. Such observables, once obtained at the physical values of quark masses, will provide important theoretical input into the designed experiments that aim to explore non-nucleonic gluons in nuclei.

\section{Challenges in multi-nucleon calculations and new developments}\label{sec:GW-noise}
\noindent
Multi-hadron correlation functions obtained from a finite sampling of QCD gauge-field configurations suffer from a sever statistical noise at late times. As is well known, while the signal to noise remains approximately constant over time in a pionic correlation function, that in a nuclear correlation function degrades exponentially in time as $\sim e^{-A(M_N-\frac{3}{2}m_{\pi})t}$, where $A$ denotes the atomic number and $m_{\pi}$ and $M_N$ are the pion and nucleon masses, respectively. Further, the nearby excitations of both the single and multi-nucleon systems pose limitations on extracting ground-state properties of the systems at moderately early times when there is still a signal. The key to the success of the program on multi-baryon systems, as mentioned in  Sec.~\ref{sec:robustness}, is to use source and/or sink operators that overlap substantially onto the ground states, extending the single-exponential region of correlation functions at early times. For example, improvement seen by the use of baryon blocks, first employed by the NPLQCD collaboration in Refs.~\cite{Beane:2006mx, Beane:2006gf}, can be understood from the intuition of a nucleus representing a collection of nucleons. Baryon blocks have proven to be surprisingly effective in studies of multi-nucleon systems even at unphysically large values of the quark masses where such assumption about the nucleus structure could not have been made a priori. Deeper understanding and further strategies have emerged during the past year to optimally extract physics from noisy correlation functions of baryons. In what follows, I divide these developments into two categories: strategies that are aimed to alleviate the excited-state contaminations in the correlation functions at early times, and those that are designed to extract physical quantities directly from the noise region at late times:

\begin{itemize}
\item{
\emph{Improvements at early times:} A variant of the variational method is Matrix Prony which was introduced in Ref.~\cite{Beane:2009kya}, and whose applicability does not rely on a symmetric matrix of correlation functions. With $p$ sets of correlation functions corresponding to distinct source or sink operators, an eigenvalue problem can be solved such that each eigenvalue obtains one of the $p$ lowest energies  of the system, assuming that the excited states beyond the $p^{th}$ state contribute negligibly to the correlation function in a given time interval. More generally, a set of linear combinations of $p$ number of correlation functions can be constructed in such a way that each combination is dominantly a single-exponential form in the $i^{th}$ energy eigenvalue with $i=1,\dots,p$. An extensive implementation of the Matrix Prony and general linear-combinations method in the single and multi-baryon correlation functions, conducted in Refs.~\cite{Beane:2009kya, Beane:2009gs, Beane:2009py} by the NPLQCD collaboration, were shown to be extremely effective in eliminating the excited-state contribution to the correlation functions at early times. An example of such improvement is shown in the left panel of Fig.~\ref{fig:prony} for the EMP of the $\Xi$ baryon. While both the smeared-point and smeared-smeared correlation functions show a constant behavior in time only for $t  \geq 17$, the appropriate linear combination of the two exhibits a plateau region in the effective mass corresponding to the ground state which starts at much earlier times, $t \geq 5$. In forming the ratios of multi-baryon to single-baryon correlation functions, the improved single-baryon correlation function can be used, providing an extended plateau region in the EMPs associated with the energy difference between the interacting and free baryons~\cite{Orginos:2015aya}. Additionally, the multi-baryon correlation functions themselves can be improved through a direct application of linear-combination methods.
\begin{figure}[t]
\centering
\includegraphics[scale=0.295]{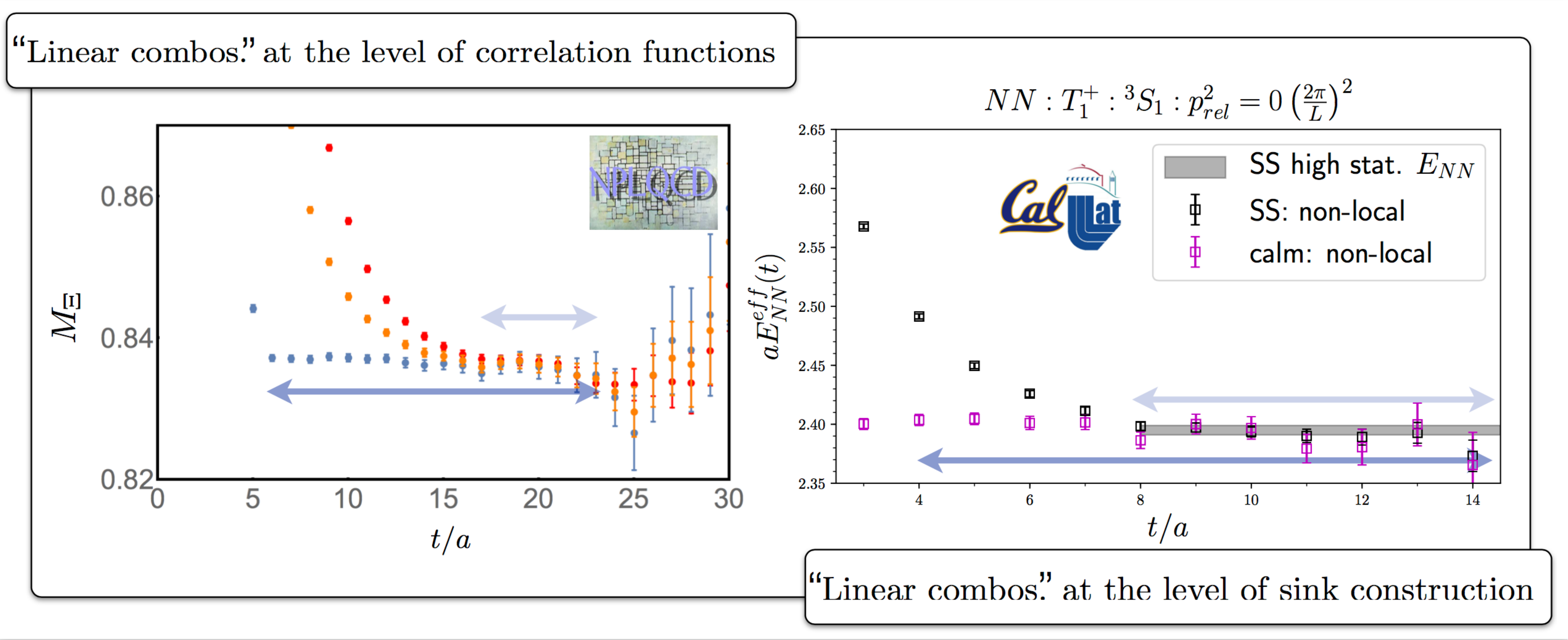}
\caption{Shown in the left is the EMP of the $\Xi$ baryon obtained from an ensemble of gauge-field configurations with $m_{\pi} \approx 450~\tt{MeV}$ and $L=32$ in lattice units~\cite{Orginos:2015aya}. A linear combination of the smeared-point (red points) and smeared-smeared (orange points) correlation functions, shown as the blue points in the figure, removes the dominant contamination from excited states and pushes the start of the plateau region to earlier times, see also Ref.~\cite{Beane:2009kya}. Masses and times are in lattice units. Shown in the right is the result of a preliminary study of the same linear combinations applied to nucleon sink operators before the two-nucleon correlation functions are formed. Similar improvement in the plateau length is seen in the two-nucleon EMPs, in particular with the use of non-local source operators. see Ref.~\cite{Berkowitz:2017smo}. The ensembles are those of Ref.~\cite{Beane:2012vq} with $m_{\pi} \approx 806~\tt{MeV}$. Permission to use the figure is granted by M. J. Savage (the left plot) and E. Berkowitz (the right plot).}
\label{fig:prony}
\end{figure}

As the sink operator for a multi-baryon system is widely chosen to be constructed from baryon blocks, different multi-baryon correlation functions can be constructed by using different sink operators for each baryon. Therefore, instead of finding appropriate linear combinations of different multi-baryon correlation functions, such combinations can be found for the single-nucleon sink operators, and these improved  operators can replace each baryon block that enters the construction of the correlation function. It is then evident that such approach should reproduce similar improvement in the early-time region of the correlation functions as that of linear-combinations method applied to the correlation functions, although  the flexibility to refine the improvement due to interactions will be lost~\cite{nplqcd}. The CalLatt collaboration implemented this approach during the past year for the two-nucleon systems, and the preliminary results, shown in the right panel of Fig.~\ref{fig:prony}, demonstrate that the expected improvement is achieved (plot taken from a proceedings to this conference ~\cite{Berkowitz:2017smo}). Even a more significant improvement is observed in the two-nucleon correlation functions that are formed from displaced sources for the nucleons, potentially pointing to the suggestion by the  collaboration that the dominant excited-state contributions to a multi-nucleon correlation function (with the present sink-source structures) arise from the excitations of a single nucleon. Displaced nucleons at the source will certainly have excitation patterns closer to that of two single nucleons, and an improvement of the nucleon operators alone should lead to a significant improvement at the level of the two-nucleon correlation function. This also means that the two-nucleon systems in the $P$-wave can benefit from a considerable improvement, as is suggested in Ref.~\cite{Berkowitz:2017smo}.
}

\item{\emph{Physics in the noise region:} An important observation made by Wagman and Savage~\cite{Wagman:2016bam, Wagman:2017xfh} in the past year reveals a clearer connection between the signal-to-noise problem of hadronic correlation functions and the sign problem. It was found that while the magnitude of the nucleon correlation function remains essentially free of noise at all times, the signal for the phase of the correlation function degrades over time. The ``ground-state'' mass associated with the magnitude and phase are found to be approximately $\frac{3}{2}m_{\pi}$ and $M_N-\frac{3}{2}m_{\pi}$, respectively. In other words, the nucleon physics is embedded in quantum fluctuations of the phase, causing  rapid changes in the sign of the real part of the correlation function, hence the connection to a sign problem. Ref.~\cite{Wagman:2016bam} provides a detailed study of the statistics of the (momentum-projected) hadronic correlation function as a function of time, and uses the acquired insight to offer alternative methods that extract the mass of the nucleon from the late-time noise-dominated region of the correlation functions~\cite{Wagman:2017xfh}. One such method is to ``freeze'' the problematic phase in the hadronic correlation functions during an initial time and let the system evolve for $t-\Delta t$ steps, free of the signal degradation, before introducing back the phase, where $t$ is the source-sink separation (see the lower-right panel of Fig.~\ref{fig:ston-23} for an illustration). This procedure suggests that a phase-reweighted correlation function can be considered 
\begin{eqnarray}
G^{\theta}(t,\Delta t)=\braket{e^{i \theta_i(t-\Delta t)}C_i(t)},
\label{eq:phase-RW}
\end{eqnarray}
\begin{figure}[t]
\centering
\includegraphics[scale=0.35]{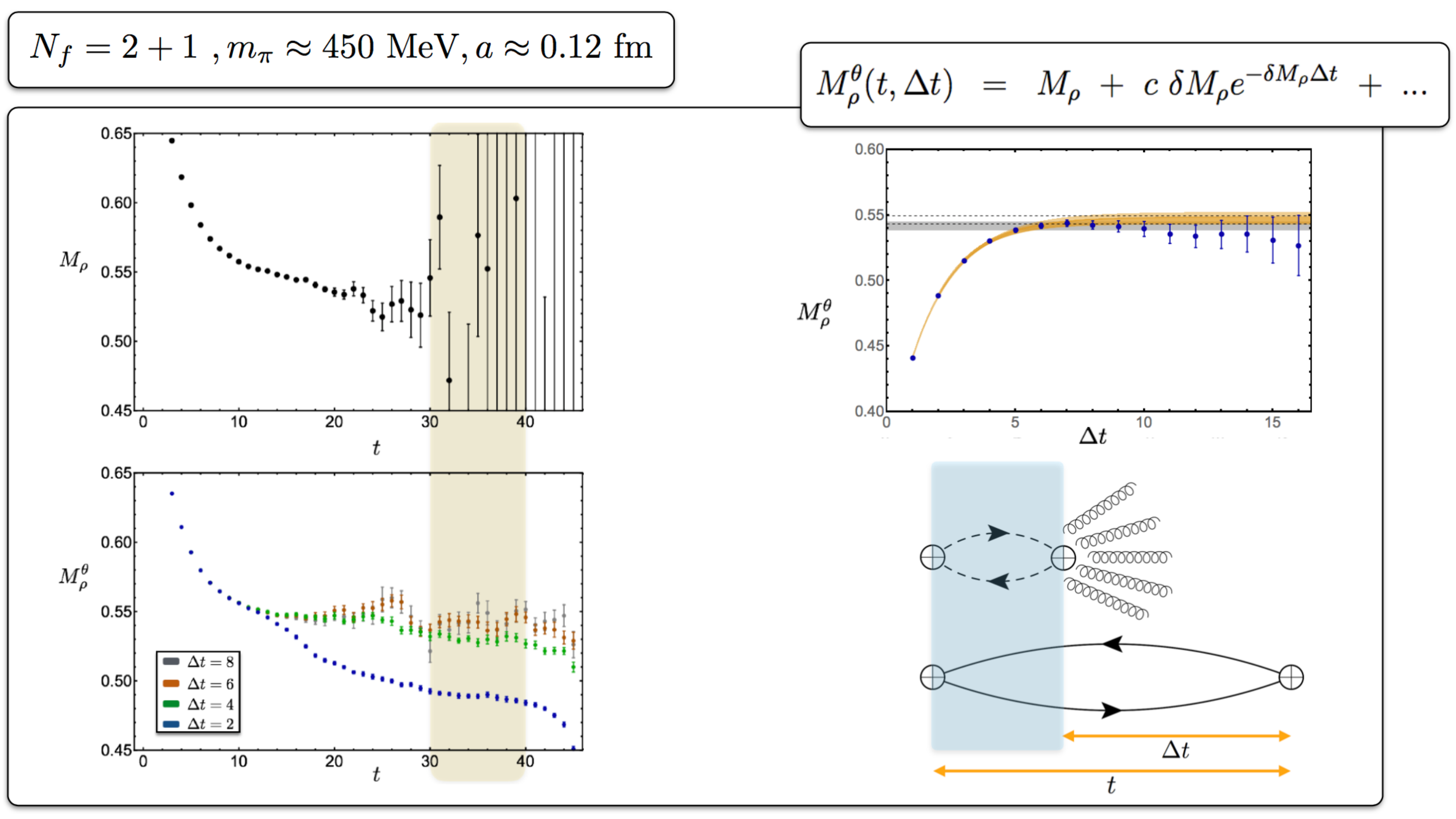}
\caption{ The EMP of the $\rho^+$, and the EMPs corresponding to the phase-reweighted correlation function with a range of fixed $\Delta t$'s are shown in the upper-left panel and lower-left panel of the figure, respectively. The highlighted interval is used for a fit to the effective mass of the phase-reweighted correlation functions. The upper-right panel shows the $\rho^+$ meson phase-reweighted effective mass for all $\Delta t \leq t$, along with a fit (the brown band) to the form shown in the figure. The region between the two dashed lines is the mass obtained from such extrapolation in $\Delta t$. The gray horizontal band corresponds to Golden-Window result of Ref.~\cite{Orginos:2015aya} obtained with four times higher statistics. An illustrative picture of the meson phase-reweighted correlation function defined in Eq. (\ref{eq:phase-RW}) is shown in the lower-right panel. It effectively includes a non-local source whose magnitude is dynamically refined for $t-\Delta t$ steps, while the phase is held fixed (shaded region) before the full system is evolved for the last $\Delta t$ steps of propagation. Quantities are given in lattice units. Plots are from Ref.~\cite{Wagman:2017xfh} and the ensembles used are those of Ref. \cite {Orginos:2015aya}. Permission to use the figure is granted by M. Wagman.}
\label{fig:ston-23}
\end{figure}
where $\theta_i(t-\Delta t)=\arg(C_i(t-\Delta t))$ and subscript $i$ denotes the gauge configuration index. Clearly, the original noisy correlation function is recovered when $\Delta t = t$. For $\Delta t \ll t$, the signal to noise in the correlation function remains essentially constant, but a considerable bias is introduced in the extracted masses. An example of the EMPs corresponding to the phase-reweighted correlation function is shown in the lower-left panel of Fig.~\ref{fig:ston-23} for the $\rho^+$ meson at a pion mass of $m_{\pi} \approx 450~\tt{MeV}$, to be compared with the EMP corresponding to the regular correlation function in the upper-left panel. As is seen, by decreasing the time during which the phase of the correlation function is suppressed (increasing $\Delta t$), the late-time noise re-appears and the original correlation function is recovered. The bias introduced by eliminating the phase of the correlation function during an initial stage of the evolution can be removed. This is done by an extrapolation of the obtained masses at multiple values of $\Delta t$ towards the $\Delta t =t$ limit, as is demonstrated in the upper-right panel of Fig.~\ref{fig:ston-23}. As is observed, the result of such extrapolation using a physically justified ansatz for its form, agrees with that obtained from the Golden-Window approach with four times higher statistics (the dashed strip compared with the brown band).

As mentioned in Ref.~\cite{Wagman:2017xfh}, the above phase-reweighting technique resembles similar techniques in the quantum Monte Carlo calculations of many-body systems in which the phase of the wave-function is suppressed until the system is sufficiently close to the ground state~\cite{Zhang:1995zz, Zhang:1996us, Wiringa:2000gb, Carlson:2014vla}, or the in lattice EFT calculations of nuclei where a sign-problem-free Hamiltonian is used at earlier stages of the evolution, after which the asymmetric perturbations away from the isospin limit are added~\cite{Lahde:2015ona}. Additionally, on the physical grounds, there appears to exist connections between the phase-reweighted correlation functions and the approximate factorization of domain-decomposed quark propagators introduced in Refs.~\cite{Ce:2016idq, Ce:2016ajy} (recent progress can be found in proceedings to this conference~\cite{Giusti:2017ksp, Ce:2016qto}). Further investigation of the phase-reweighting technique for LQCD correlation functions, in particular in the mesonic sector and in nuclei, is underway, and may promise an effective way to obtain results from noisy correlation functions in the multi-hadron sector.
}
\end{itemize}

\section{Summary}\label{sec:summary}
\noindent
Lattice quantum chromodynamics (LQCD) presents the promise of reducing some of the outstanding theoretical uncertainties in important nuclear quantities, from low-energy reaction rates for astrophysics to nuclear modifications to the Standard Model (SM) and beyond the SM processes in nature. Since the introduction of LQCD to nuclear physics, significant progress has been made in obtaining the lowest-lying spectra of few-baryon systems~\cite{Beane:2012vq, Yamazaki:2012hi, Orginos:2015aya, Yamazaki:2015asa}, along with their scattering amplitudes and interactions~\cite{Beane:2006gf,Nemura:2008sp,Beane:2009py,Beane:2010hg,Beane:2011zpa,Beane:2011iw,Inoue:2011ai,Beane:2012ey,Yamada:2015cra,Berkowitz:2015eaa, Miyamoto:2017tjs, Gongyo:2017fjb}. Most recently, these studies have been expanded such that nuclear matrix elements relevant for some of the primary reaction processes in light nuclear systems, such as $np \to d\gamma$, $pp$ fusion, tritium beta decay and the di-neutron's double beta decay~\cite{Beane:2015yha, Savage:2016kon, Shanahan:2017bgi, Tiburzi:2017iux}, could be obtained. Additionally, developments in nuclear matrix elements from LQCD has enabled studies of the low-energy structure properties of light nuclei as probed by the quark and gluon probes~\cite{Savage:2016kon, Winter:2017bfs}. The validity of different approaches that are practiced by the community, and their associated systematics, have been the subject of careful investigations, and the community is engaged in vigorous discussions to establish the robustness of the multi-nucleon calculations to date~\cite{Beane:2010em, Walker-Loud:2014iea, Yamazaki:2015nka, Savage:2016egr, Yamazaki:2017gjl, Iritani:2016jie, Iritani:2017rlk, Beane:2017edf, Wagman:2017tmp, Yamazaki:2017euu, Yamazaki:2017jfh}. The complexity of multi-baryon correlation functions has constrained the application of a larger basis of interpolating operators that can raise the confidence in the extracted finite-volume spectra but progress is underway. Further, the exponential degradation of the signal to noise in systems involving multiple hadrons remains an issue.  Nonetheless, new ideas have been suggested in recent years to alleviate these problems, and the results presented previously and in this conference have proven to be promising~\cite{Beane:2009kya, Berkowitz:2017smo, Wagman:2016bam, Wagman:2017xfh}. As the community moves towards the \emph{Exascale} computing era~\cite{ExascaleNP}, and with the advent of new algorithms and novel ideas along the way, the LQCD and QED calculations with physical masses of the light quarks, at larger lattice volumes and with multiple lattice spacings, may become a reality in upcoming years. Such SM input in the few-baryon sector with controlled uncertainties, when combined with proper effective fields theories and advanced quantum many-body calculations, will go a long way to advance our understanding of nuclei and dense matter, and our interpretations of experimental searches for new physics using nuclear targets.

\section*{Acknowledgment}
I would like to thank the organizing committee of ``\emph{the 35th International Symposium on Lattice Field Theory}'' for a successful scientific meeting in beautiful Granada, Spain. I would like to also thank the members of LQCD Nuclear Physics community, in particular S. Aoki, E. Berkowitz, T. Doi, T. Iritani, H. Nemura, M. Savage, P. Shanahan, M. Wagman and T. Yamazaki for sharing the results presented in this review. Special thanks to the members of the NPLQCD collaboration, S.  Beane, E. Chang, W. Detmold, K. Orginos, A, Parreno, M. Savage, P. Shanahan, B. Tiburzi, M. Wagman and F. Winter,  for valuable comments, inspiring discussions and continuous scientific collaboration on the topics of this review. This work was supported by the U.S. Department of Energy under grant number DE-SC0011090, and by the Maryland Center for Fundamental Physics, College Park, MD.

\bibliography{lattice2017}

\end{document}